\documentclass[fleqn]{article}
\usepackage{amsmath,amssymb}
\usepackage{epsfig,graphicx}

\renewcommand{\theequation}{\arabic{section}.\arabic{equation}}

\catcode`@=11 \@addtoreset{equation}{section} \catcode`@=12

\begin{document}
\title{\vskip-1.7cm \bf  Aspects of Nonlocality in Quantum Field Theory, Quantum Gravity and Cosmology}
\date{}
\author{A.O.Barvinsky}
\maketitle \hspace{-8mm} {\qquad\em
Theory Department, Lebedev
Physics Institute, Leninsky Prospect 53, Moscow 119991, Russia}

\begin{abstract}
This paper contains a collection of essays on nonlocal phenomena in quantum field theory, gravity and cosmology. Mechanisms of nonlocal contributions to the quantum effective action are discussed within the covariant perturbation expansion in field strengths and spacetime curvatures and the nonperturbative method based on the late time asymptotics of the heat kernel. Euclidean version of the Schwinger-Keldysh technique for quantum expectation values is presented as a special rule of obtaining the nonlocal effective equations of motion for the mean quantum field from the Euclidean effective action. This rule is applied to a new model of ghost free nonlocal cosmology which can generate the de Sitter stage of cosmological evolution at an arbitrary value of $\varLambda$ -- a model of dark energy with its scale played by the dynamical variable that can be fixed by a kind of a scaling symmetry breaking mechanism. This model is shown to interpolate between the superhorizon phase of gravity theory mediated by a scalar mode and the short distance general relativistic limit in a special frame which is related by a nonlocal conformal transformation to the original metric. The role of compactness and regularity of spacetime in the Euclidean version of the Schwinger-Keldysh technique is discussed.
\end{abstract}

\maketitle

\tableofcontents

\section{Introduction}
\hspace{\parindent}It is well known that the corner stone laid in the foundations of new physical models is the principle of locality, when the equations of motion for fundamental variables of the theory are local in time and space and admit a well-posed initial value problem. This setup allows one to build the Hamiltonian and Lagrangian formalisms, perform canonical quantization with the notion of instant quantum state of the system, establish its causality and so on. In fact this starting principle continues dominating high energy physics even in the description of extended objects, because locality still persists in the fundamental spacetime, like 2D worldsheet in string theory, as opposed to effectively observable spacetime composed of zero modes of string variables. Attempts to overstep the locality principle at foundation level, starting with one of the old approaches \cite{Efimov}, continue, but the number of open issues and ambiguities is such that they do not yet form a well recognized avenue towards unified theory of interactions.

On the other hand, nonlocal phenomena play a very important role in classical and quantum physics. Needless to say that any solution of local equations of motion bears a nonlocal dependence both in space and time on the initial data. Moreover, at the quantum level even the equations of motion become nonlocal when they are derived from local Heisenberg equations for {\em effective mean} field -- expectation value of the quantum variable or its matrix element between prescribed initial and final quantum states. In this case no ambiguities in the definition of nonlocal objects and their boundary conditions takes place, and everything can be directly derived from the physical setup of the problem. This is the approach which we will stick to in the discussion of the nonlocal equations of motion and the nonlocal effective action which generates them.

This paper contains a collection of essays on nonlocal phenomena in quantum field theory, gravity and cosmology. We start with the discussion of mechanisms for nonlocal contributions to the quantum effective action, which in contrast to low-energy vacuum polarization effects by massive quantum fields characterize high-energy asymptotics in
massive theories or infrared behavior in theories of massless
fields. In Sect.2 we begin with the heat kernel method of Schwinger and DeWitt \cite{DeWitt,PhysRep} as a basis of local expansion in massive theories, show breakdown of this expansion in the massless limit and develop alternative approximation techniques -- nonlocal covariant expansion in field strengths (spacetime and fibre bundle curvatures in the gravity theory context) and nonperturbative method associated with the late time asymptotics of the heat kernel. Important part of this discussion is the relation between the formal expansion in Euclidean spacetime and concrete problems in the physical Lorentzian signature spacetime for quantum expectation values. This relation, which can be called the Euclidean version of the Schwinger-Keldysh technique \cite{SchwKeld,Vilkovisky_lectures}, serves as a guiding principle for the treatment of nonlocalities in effective equations of nonlocal cosmology in Sects.3 and 4.

Irrespective of the origin of nonlocality in equations of motion, the interest in nonlocal gravity and cosmology is strongly motivated now by the attempts of resolving the cosmological acceleration (dark energy) and dark matter problems \cite{acceleration}. Deeply infrared nature of these phenomena and the massless nature of the graviton in general relativity (GR) imply that their mechanism can be essentially nonlocal, and it is up to a relevant nonlocal long-distance modification of the Einstein theory to explain them. This modification can successfully compete with popular phenomenological mechanisms of infrared modifications, induced, say, by braneworld scenarios with extra dimensions \cite{RS-GRS,DGP,Deffayet} or other models \cite{MOND,f(R)}. However, the search for such a modification is bounded by strong restrictions following from the requirement of consistency -- classical and quantum stability, unitarity and causality -- and the necessity to fit observations.

We discuss these issues in Sects.3 and 4 and, in particular, suggest an effective nonlocal model of gravity theory based on the idea of the scale dependent gravitational coupling \cite{AHDDG,covnonloc}. It is free of ghosts and capable of generating the de Sitter stage of the cosmological evolution at {\em any} value of the cosmological constant scale -- a possible route to the solution of cosmic coincidence problem by means of replacement of the numerical scale by a dynamical variable with its value selected by a kind of a scaling symmetry breaking mechanism. Important ingredient of this model is the Euclidean version of the Schwinger-Keldysh technique which, when extended to the de Sitter spacetime setup \cite{HiguchiMarolfMorrison,Tanaka} gives unambiguous rules for nonlocal effective equations for the mean metric field. Basing on these rules we prove stability of the de Sitter background, obtain massless gravitons as free propagating modes on top of it and get retarded gravitational potentials of matter sources. In a special conformal frame these potentials show that the theory interpolates between the short distance GR phase and the the superhorizon scale at which the interaction is mediated by a scalar conformal mode.

\section{Quantum effective action as a source of nonlocality}
\hspace{\parindent} The origin of nonlocal terms from quantization of local theories can be easily demonstrated in the one-loop approximation by the heat kernel method. One-loop quantum effective action for a theory with the classical action $S[\,\varphi\,]$ and the wave operator $F(\nabla)\,\delta(x,y)
=\delta^2 S[\varphi]/\delta\varphi(x)\delta\varphi(y)$ has a proper time integral representation
    \begin{eqnarray}
     &&\varGamma=\frac12\,{\mathrm{Tr}} \,\ln F(\nabla)=
     -\frac{1}{2}\,
     \int_{0}^{\infty }\frac{ds}{s}\:
     {\rm Tr} K(s)\,,                   \label{1.10}\\
     &&{\rm Tr} K(s)
     =\int dx\, K(s|\,x,x).              \label{1.11}\\
     &&K(s|\,x,y)=e^{sF(\nabla)}\,\delta(x,y).   \label{1.7}
    \end{eqnarray}
in terms of the functional trace of the heat kernel $K(s|\,x,y)$ which solves the heat equation with a unit initial condition at $s=0$
    \begin{equation}
     \frac{\partial K(s)}{\partial s}
     =F(\nabla)\,K(s).                        \label{1.8}
    \end{equation}
Its expansion in powers of $s$ underlies the technique of the local Schwinger-DeWitt expansion for massive fields \cite{DeWitt,PhysRep}, whereas its late time limit $s\to\infty$ is responsible for infrared nonlocal effective action of massless models \cite{CPTI,CPTII,CPTIII,asymp}. In this section we show how the breakdown of local expansion calls for alternative nonlocal approximation schemes for the heat kernel and effective action.

\subsection{Heat kernel method and breakdown of local expansion}
\hspace{\parindent}For brevity we start with the case of the quantized field without spin indices in flat spacetime with the metric $g_{\mu\nu}=\delta_{\mu\nu}$ and later generalize it to curved spacetime including gravity. The efficiency of the above proper time representation is based on the fact that for theories with the second-order operator of the form
    \begin{equation}
     F(\nabla)=\Box-V(x)-m^2,\quad
     \Box=g^{\mu\nu}\nabla_\mu\nabla_\nu      \label{1.6}
    \end{equation}
this heat kernel has a small $s$ expansion \cite{DeWitt,Gilkey,PhysRep,Avramidi,Vasilevich}
    \begin{eqnarray}
     && K(s|\,x,y)=\frac1{(4\pi s)^{d/2}}\,
     e^{-|x-y|^2/4s-m^2s}\,
     \Omega(s|\,x,y),\quad
     \Omega (s|\,x,y)=\sum\limits_{n=0}^{\infty}
     a_{n}(x,y)\,s^{n},         \label{ansatz}
    \end{eqnarray}
where $d$ is the spacetime dimensionality. This semiclassical
ansatz for the heat kernel guarantees that at $s\to 0$ it tends to the delta-function $\delta(x,y)$ and contains all nontrivial information about the potential $V(x)$ in the function
$\Omega(s|\,x,y)$ which is analytic at $s=0$.

The coefficients of this expansion play a very important role in quantum field theory and have the name of HAMIDEW coefficients that was coined by G.Gibbons \cite{Gibbons} to praise  joint efforts of mathematicians and physicists in heat kernel theory and its implications. The heat equation gives a set of recurrent equations for $a_{n}(x,y)$, which can be solved in a closed form for their coincidence limits at $y=x$. The result for $a_{n}(x,x)$ turns out to be local in terms of the potential $V(x)$ and its multiple derivatives. For the operator (\ref{1.6}) in flat spacetime the first few of them read
    \begin{eqnarray}
     && a_0(x,x)=1, \quad a_1(x,x)=-V(x), \quad a_2(x,x)=\frac12\,
     V^2(x)+\frac16\, \Box V(x),            \label{2.3}
    \end{eqnarray}
The dimensionality of $a_{n}(x,x)$ in units of
inverse length grows with $n$ and is comprised of the powers of the potential $V(x)$ and its derivatives.

Substituting the expansion (\ref{ansatz}) into
(\ref{1.10}) we obtain the one-loop effective action in the form of the asymptotic $1/m^2$ series
\cite{DeWitt,PhysRep,CPTII}
    \begin{eqnarray}
     &&\frac12\,{{\rm Tr}}\ln\,F(\nabla)
     =\varGamma_{\mathrm{div}}
     +\varGamma_{\log }-\frac{1}{2}
     \left( \frac{m^{2}}{4\pi}\right)^{d/2}
     \int dx\,\sum\limits_{n=d/2+1}^{\infty} \,\Gamma(n\!-\!d/2)
     \frac{a_{n}(x,x)}{(m^{2})^{n}}\,.  \label{2.6}
    \end{eqnarray}
The first $d/2$ integrals (we assume that $d$ is even) are
divergent at the lower limit and generate ultraviolet divergences $\varGamma_{\mathrm{div}}$ accompanied by the
logarithmic term $\varGamma_{\log }$. In dimensional regularization they read
    \begin{eqnarray}
    &&\varGamma_\mathrm{div}\!=\!\frac{1}{2(4\pi )^{d/2}}\int
    dx\,\sum\limits_{n=0}^{d/2}
    \left[-\frac{1}{d/2-\omega}
    -\psi\Big(
    \frac{d}{2}-n+1\Big)\right]
    \frac{(-m^2)^{d/2-n}}{(d/2
    -n)!}\,a_{n}(x,x),       \label{2.6a}\\
    &&\varGamma_{\log }=\frac{1}{2(4\pi )^{d/2}}\int
    dx\,\sum\limits_{n=0}^{d/2}\,
    \frac{(-m^{2})^{{d/2}-n}}{\left({d/2}
    -n\right) !}
    \ln \frac{m^{2}}{\mu^{2}}\,a_{n}(x,x),       \label{2.6b}
    \end{eqnarray}
where $2\omega\to d$ and $\mu^2$ is the normalization mass parameter reflecting the renormalization ambiguity. All these equations can be generalized to the case of the generic spin-tensor field in curved spacetime when the operator $F(\nabla)$, its potential, heat kernel $K(s)$ and coefficients $a_{n}(x,y)$ acquire matrix structure and $\nabla_\mu$ become covariant derivatives. Then the spacetime integration also acquires Riemann measure $g^{1/2}$ and $a_{n}(x,x)$ in (\ref{2.6b}) imply taking the trace of the matrix valued coefficients $a_{n}(x,y)$. The latter in addition to (\ref{2.3}) get terms built of spacetime curvature, fibre bundle curvature ${\cal R}_{\mu\nu}=[\nabla_\mu,\nabla_\nu]$ and their covariant derivatives.

This local expansion makes sense only when the terms of the asymptotic series rapidly decrease with the growth of
$n$, that is when the mass parameter $m$ is large compared
\emph{both} to the derivatives of the potential and the
potential itself. In the presence of the gravitational field these restrictions include also the smallness of spacetime and fibre bundle curvatures and their covariant derivatives compared to the mass parameter. For large field strengths or rapidly varying fields the
Schwinger-DeWitt expansion becomes inapplicable and completely
blows up in the massless limit $m\rightarrow 0$ when the proper time integration becomes divergent at $s\to\infty$ in various terms of this expansion. Below we consider two perturbation methods which extend the Schwinger-DeWitt technique to the class of massless models and the nonperturbative technique based on the late-time asymptotics of the heat kernel.

The first method may be called the modified Schwinger-DeWitt expansion, which corresponds to resummation of the original expansion leading to exponentiation of $-sV(x)$ in (\ref{ansatz}). When the potential is positive-definite (which we shall assume here) it can play the role of the cutoff factor similar to the mass term in the integral (\ref{2.6}). The relevant ansatz instead of (\ref{ansatz}) reads
    \begin{eqnarray}
     &&K(s|\,x,y)=\frac1{(4\pi s)^{d/2}}\,
     e^{-|x-y|^2/4s-sV(x)}\;
     \tilde{\Omega}(s\,|\,x,y),\quad
     \tilde{\Omega} (s|\,x,y)
     =\sum\limits_{n=0}^{\infty }
     \tilde{a}_{n}(x,y)\,s^{n}.           \label{2.9}
    \end{eqnarray}
Here the new function $\tilde{\Omega}(s\,|\,x,y)$ has the expansion in $s$ with the modified Schwinger-DeWitt coefficients $\tilde{a}_{n}(x,x)$ which contain only powers of the differentiated potential and vanish for $\nabla
V=0$.

Now the proper time integral even for $m^2=0$ has
an the infrared cutoff  at $s\sim 1/V(x)$ and in this case the
effective action is similar to (\ref{2.6}), where $m^{2}$ is
replaced by $V(x)$ and $a_{n}(x,x)/m^{2n}$ by $\tilde{a}_{n}(x,x)/V^n(x)$. In particular, the ultraviolet divergences are given by the massless limit of $\varGamma_\mathrm{div}$ and the logarithmic part gives rise (in $d=4$) to the Coleman-Weinberg term
$\varGamma_{\log}\rightarrow\varGamma_\mathrm{CW}+O(\nabla V)$,
    \begin{eqnarray}
     \varGamma_\mathrm{CW}=\frac{1}{64\pi^2}
     \int dx\, V^2(x)\,
     \ln \frac{V(x)}{\mu ^{2}},                 \label{2.11}
    \end{eqnarray}
corrected by the contribution due to the derivatives of $V(x)$. $\varGamma_\mathrm{CW}$ here is the spacetime integral of the Coleman-Weinberg effective potential which in the
$\varphi^4$-model with $V(\varphi)\sim\varphi^2$reduces to the well-known expression $\varphi^4\ln (\varphi^2/\mu^{2})/64\pi
^{2}$.

The modified Schwinger-DeWitt expansion runs in the derivatives of $V$ rather than powers of $V$ itself because the typical structures entering $\tilde{a}_{n}(x,x)$ are represented by $m$ derivatives acting in different ways on the product of $j$
potentials, $\nabla ^{m}V^{j}(x)$. Here $m+2j=2n$ and $m\geq j$ because every $V$ should be differentiated at least once. Such an expansion is efficient if the potential is slowly varying in units of the potential itself $\nabla\nabla/V\ll 1$. When $V(x)$ is bounded from below by a large positive constant this condition can be easily satisfied throughout the whole spacetime. But this case is uninteresting because it reproduces the original Schwinger-DeWitt expansion with $m^2$ playing the role of this bound. More interesting is the case of the asymptotically empty spacetime when the potential and its derivatives fall off to zero by the power law $V(x)\sim |x|^{-p}$, $\nabla ^{m}V(x)\sim |x|^{-p-m}$, $|x|\rightarrow \infty$, for some positive $p$. For such a potential terms of the perturbation series behave as
    \begin{equation}
    \frac{\tilde{a}_{n}(x,x)}{V^{n}(x)}\sim
    \sum\limits_{j=1}^{[2n/3]}
    \frac{\nabla ^{2n-2j}V^{j}}{V^n}\sim
    \sum\limits_{j=1}^{[2n/3]}|x|^{(p-2)(n-j)}   \label{exppar1}
    \end{equation}
and decrease with $n$ only if $p<2$. Thus, this expansion makes sense only for slowly decreasing potentials with $p<2$. However, in this case the potential $V(x)$ is not integrable
over the whole spacetime $\left( \int dx\,V(x)=\infty \right)$ and,
moreover, even the operation $(1/\Box )V(x)$ is not well
defined\footnote{For the convergence of the integral in
$\left(1/\Box \right) V$ the potential $V(x)$ should fall off at least as $1/|x|^{3}$ in any spacetime dimension \cite{CPTII}.}.
Therefore, the above restriction is too strong to account for
interesting physical problems in which the parameter $p$ typically coincides with the spacetime dimensionality $d$. In addition, the modified asymptotic expansion is completely local and does not allow one to capture nonlocal terms of effective action.

Thus an alternative technique is needed to obtain the late-time contribution to the proper-time integral and, in particular, to understand whether and when this integral exists at all in
massless theories. The answer to this question lies in the
late-time asymptotics of the heat kernel at $s\rightarrow\infty$ which can be perturbatively analyzed within the covariant perturbation theory of \cite{CPTI,CPTII,CPTIII,asymp}.

\subsection{Covariant perturbation theory and the Euclidean version of Schwinger-Keldysh technique}
\hspace{\parindent} In the covariant perturbation theory the full potential $V(x)$ is treated as a perturbation and the solution of the heat equation is found as a series in its powers. From the
viewpoint of the Schwinger-DeWitt expansion it corresponds to an
infinite resummation of all terms with a given power of the
potential and arbitrary number of derivatives. The result reads as
    \begin{eqnarray}
    &&{{\rm Tr}}K(s)\equiv \int dx\,K(s|\,x,x)=
    \sum\limits_{n=0}^{\infty }
    {{\rm Tr}}K_{n}(s),           \label{CPT1}\\
    &&{{\rm Tr}}K_{n}(s)=\int
    dx_{1}dx_{2}...dx_{n}\,
    K_{n}(s|\,x_{1},x_{2},...x_{n})
    \,V(x_{1})V(x_{2})...V(x_{n}),
    \end{eqnarray}
with the nonlocal form factors $K_{n}(s|\,x_{1},x_{2},...x_{n})$ which
were explicitly obtained in \cite{CPTI,CPTII,CPTIII} up to $n=3$ inclusive. In the presence of gauge fields and gravity this expansion can be generalized by including the spacetime $R^\mu_{\;\nu\alpha\beta}$ and fibre bundle $\mathcal{R}_{\mu\nu}$ curvatures in the full set of perturbatively treated field strengths,
$V\rightarrow \mathcal{R}=(V,\mathcal{R}_{\mu\nu},R_{\mu\nu})$ and covariantizing the corresponding nonlocal form factors.

The expansion related to (\ref{CPT1}) also exists for the one-loop effective action $\varGamma$
    \begin{eqnarray}
    \varGamma=\sum\limits_{n=0}^{\infty }\int
    dx_{1}dx_{2}...dx_{n}\,
    \varGamma_{n}(s|\,x_{1},x_{2},...x_{n})
    \,V(x_{1})V(x_{2})...V(x_{n}),
    \end{eqnarray}
with the relevant form factors
    \begin{eqnarray} \varGamma_{n}(x_{1},x_{2},...x_{n})=-\frac12\int_0^\infty \frac{ds}s\, K_{n}(s|\,x_{1},x_{2},...x_{n})\,.
    \end{eqnarray}
It was shown \cite{CPTII} that at $s\rightarrow \infty$  ${{\rm Tr}}K_{n}(s)=O(s^{1-d/2})$, $n\geq 1$, and, therefore in spacetime dimension $d\geq 3$ these integrals are infrared convergent. In one and two dimensions this expansion for $\varGamma$ does not exist except for the special case of the massless theory in curved two-dimensional spacetime, when it reproduces the Polyakov action \cite{CPTII}, $\varGamma_\mathrm{Polyakov}\sim\int d^2 x\,g^{1/2}\,R(1/\Box) R$, which originally was obtained by integrating the conformal anomaly \cite{Polyakov}. On the contrary, in $d=4$ and beyond this expansion is well defined modulo ultraviolet divergences. In four dimensions, in particular, the lowest order form factors read \cite{CPTI,CPTII,CPTIII,spectral}
    \begin{eqnarray}
    &&\varGamma_{2}(x_{1},x_{2})=\frac1{64\pi^2}\,
    \ln\left(\frac{-\Box_1}{\mu^2}
    \right)\delta(x_1-x_2),              \label{2point}\\
    &&\varGamma_{3}(x_{1},x_{2},x_{3})=
    -\frac1{32\pi^2}\int\limits_{\alpha_i\geq 0}\frac{d^3\alpha\,
    \delta\big(1-\sum_{i=1}^3\alpha_i\big)}
    {\alpha_2\alpha_3\Box_1+\alpha_3\alpha_1\Box_2
    +\alpha_1\alpha_2\Box_3}\,
    \delta(x_1-x_2)\delta(x_1-x_3),       \label{3point}
    \end{eqnarray}
where $\mu^2$ is the normalization parameter absorbing the logarithmic divergences and the subscript of $\Box_i$ indicates on which of the coordinates $x_i$ it is acting.

Thus far this expansion was considered formally by treating $\Box$-operator arguments as numerical parameters. This is justified in Euclidean signature asymptotically-flat spacetime when the operator $\Box$ is subject to zero boundary conditions at infinity and, therefore, positive definite. For this reason we will call the above form factors and their effective action the {\em Euclidean} ones and, if necessary, label them by the subscript (or superscript) $E$). The non-polynomial and, moreover, nonanalytic dependence on them poses the question of how to interpret the resulting nonlocalities in Lorentzian signature spacetime where the operator $\Box$ is indefinite and has infinite set of zero modes -- propagating solutions of Klein-Gordon equation. Covariant perturbation theory knows the answer to this question -- what kind of boundary conditions these nonlocalities should satisfy depending on the physical setting of the problem.

Basically there are two problems known in quantum field theory -- scattering problem with the two IN and OUT vacuum states and the IN-IN problem for expectation values in the initial quantum state $|\,{\rm in}\rangle$. They both are characterized by correlation functions beginning with the lowest order one -- the mean field $\phi(x)= \phi_{F}(x)$, $\phi_{IN}(x)$,
    \begin{eqnarray}
    &&\phi_{F}(x)=
    \frac{\langle {\rm out}\,|\,\hat \phi(x)\,|\,{\rm in}\rangle}{\langle {\rm out}\,|\,{\rm in}\rangle},\quad
    \phi_{IN}(x)=
    \langle {\rm in}\,|\,\hat \phi(x)\,
    |\,{\rm in}\rangle.                         \label{meanfield}
    \end{eqnarray}
The mean field satisfies the effective equations containing the classical term and quantum corrections in the form of the radiation current $J(x)$,
    \begin{eqnarray}
    &&\frac{\delta S}{\delta\phi(x)}+J(x)=0,\\
     &&J(x)=\frac{i}2\int dy_1 dy_2\,\frac{\delta^3 S[\,\phi\,]}
     {\delta\phi(x)\delta\phi(y_1)\delta\phi(y_2)}\,
     G(y_1,y_2)
     +{\rm multiloop\; orders}.         \label{1loopradcur}
    \end{eqnarray}

The radiation currents $J= J_F$, $J_{IN}$ for these two problems are built by different diagrammatic techniques -- the usual Feynman one for a scattering case (this explains the notation $\phi_n$) and the Schwinger-Keldysh technique \cite{SchwKeld,SchwKeld2,CPTI,Vilkovisky_lectures} for $\phi_{IN}(x)$. In the one-loop approximation it has the above expression with two different Green's functions $G(x,y)= G_{F}(x,y)$, $G_{IN}(x,y)$,
    \begin{eqnarray}
    &&G_{F}(x,y)=
    \frac{\langle {\rm out}\,|\,T\,\big(\hat\phi(x)\,
    \hat\phi(y)\big)\,|\,{\rm in}\rangle}{\langle{\rm out}\,|\,{\rm in}\rangle},\quad
    G_{IN}(x,y)=
    \langle {\rm in}\,|\,T\,\big(\hat\phi(x)\,\hat \phi(y)\big)\,|\,{\rm in}\rangle
    \end{eqnarray}

In the covariant perturbation theory of the above type these currents have the expansion
    \begin{equation}
    J_{F,IN}(x)=\sum\limits_{n=0}^{\infty }\int
    dy_{1}dy_{2}...dy_{n}\,
    \varGamma_{n}^{F,IN}(x\,|\,y_{1},y_{2},...y_{n})
    \,V(y_{1})V(y_{2})...V(y_{n}),  \label{Jn}
    \end{equation}
with the nonlocal coefficients -- formfactors $\varGamma_{n}^{F,IN}(x\,|\,y_{1},y_{2},...y_{n})$. A well-known statement is that Feynman form factors can be obtained by Wick rotation from the relevant form factors of the Euclidean field theory. In our case this is $\varGamma_{n}^{F}(x\,|\,y_{1},y_{2},...y_{n})=
(n+1)\,{V}_\phi(\nabla_x)\varGamma_{n}^{E}(x,y_{1},y_{2},...y_{n})|_{\,\rm Wick}$, where ${V}_\phi(\nabla)$ is a local vertex operator, $\delta V(x)/\delta\phi(y)={V}_\phi(\nabla)\delta(x-y)$. A similar statement regarding the IN-IN form factors can be called the Euclidean version of the Schwinger-Keldysh technique which guarantees causality of the effective equations for $\phi_{IN}(x)$ -- the fact that they contain only quantities belonging to the past of the point $x$, the argument of the mean field $\phi_{IN}(x)$.

This property manifesting unitarity and locality of the underlying quantum field theory can, in principle, be realized in different ways. However, the Schwinger-Keldysh formalism dictates one concrete way of obtaining $\varGamma_{n}^{F}(x\,|\,y_{1},y_{2},...y_{n})$. When the IN-state $|\,in\rangle$ is a {\em Poincare-invariant vacuum} associated with the past asymptotically flat infinity, then $\varGamma_{n}^{F}(x\,|\,y_{1},y_{2},...y_{n})$ follows from the Euclidean form factor by a formal transition to the Lorentzian signature spacetime and taking in all nonlocalities the retardation rule for all points $y_i$ relative to the observation point $x$. Technically, the proof of this statement, which was done to the first order of perturbation theory in \cite{Hartle-Horowitz} and to all orders of the curvature expansion in \cite{CPTI}, runs in the momentum space representation
    \begin{eqnarray}
    &&\varGamma_n^{F,IN}(x\,|\,y_1,...y_n)=\frac1{(2\pi)^{4n}}\int d^4k_1...d^4k_n\,
    \exp\left[i\sum\limits_{l=1}^n k_l(x-y_l)\right]\,\gamma_{n}^{F,IN}(x\,|\,k_1,...k_n).
    \end{eqnarray}
The Fourier images of the IN-OUT and IN-IN form factors turn out to be the limiting values on the real axis in the complex plane plane of the momentum arguments $z^0_l=k^0_l+ik^4_l$ (for all $k_l^0$, $l=1,...n$) of the one analytic function $f_{n}(x\,|\,z_1,...z_n)$ of $z^0_l$, which is analytic outside of the real axes of $z^0_l$,
    \begin{eqnarray}
    &&\gamma_{n}^{F}(x\,|\,k_1,...k_n)=f_{n}(x\,|\,z_1,...
    z_n)\,\big|_{\,z^0=k^0(1+i\varepsilon)},\\
    &&\gamma_{n}^{IN}(x\,|\,k_1,...k_n)=f_{n}(x\,|\,z_1,...
    z_n)\,\big|_{\,z^0=k^0+i\varepsilon},\quad\varepsilon\to+0.
    \end{eqnarray}
On the contrary, the Euclidean theory form factor $\gamma_{n}^{E}(x\,|\,\tilde k_1,...\tilde k_n)$ in the Euclidean momentum space of $\tilde k=(k^4,{\bf k})$ represents the value of this function in the ``middle" of its analyticity domain -- at the imaginary axis of $z^0$-arguments,
$\gamma_{n}^{E}(x\,|\,\tilde k_1,...\tilde k_n)=f_{n}(x\,|\,z_1,...z_n)\,\big|_{\,z^0=ik^4}$. Therefore, the IN-IN form factor follows from its Euclidean counterpart by the retardation rule $\gamma_{n}^{IN}(x\,|\,k_1,...k_n)=\gamma_{n}^{E}(x\,|\,\tilde k_1,...\tilde k_n)\,|_{\,k^4=-i(k^0+i\varepsilon)}$, whereas the Feynman form factors arises from the familiar Wick rotation $k^4=-ik^0(1+i\varepsilon)$. As the result, analyticity of $f_{n}(x\,|\,z_1,...z_n)$ in the upper half of the complex plane of $z_l^0$ implies that the IN-IN formfactors have {\em retarded} nonlocality
    \begin{eqnarray}
    &&\varGamma_n^{IN}(x\,|\,y_1,...y_n)\nonumber\\
    &&\qquad\quad=\frac1{(2\pi)^{4n}}\int d^4k_1...d^4k_n\,
    e^{i\sum\limits_{l=1}^n k_l((x-y_l)}\,F_{n}(x\,|\,z_1,...z_n)\,\Big|_{\,z^0=k^0+i\varepsilon}
    =0,\quad x^0-y^0<0.
    \end{eqnarray}

Thus, finally the Euclidean version of the Schwinger-Keldysh technique states that
    \begin{eqnarray}
    &&J_{IN}(x)=
    \left.\frac{\delta\varGamma_{E}^{\rm loop}}
    {\delta\phi(x)}\,\right|_{\;++++\,\to\,-+++}^{\;\rm retarded}\;.
    \end{eqnarray}
Technically this retardation rule can be implemented by writing down for the form factors their spectral representations in terms of the mass integrals of massive Green's functions and then taking these Green's functions as the retarded ones. In particular, the radiation currents induced by the 2-point and 3-point form factors (\ref{2point})-(\ref{3point}) read
    \begin{eqnarray}
    &&J^{IN}_1(x)=\frac1{32\pi^2}{V}_\phi(\nabla)
    \int\limits_0^\infty dm^2\,\left(\frac1{m^2+\mu^2}-\frac1{m^2-\Box}\,\Big|_{\,\rm ret}\right)\,V(x),\\
    &&J^{IN}_2(x)=\frac3{32\pi^2}{V}_\phi(\nabla_x)
    \int\limits_0^\infty dm^2_1\,dm_2^2\,dm_3^2\;\rho(m_1,m_2,m_3) \left.\frac1{m^2_1-\Box_x}\;\right|_{\,\rm ret}\nonumber\\
    &&\qquad\qquad\qquad\qquad\times\left[\,\frac1
    {m^2_2-\Box_2}\;\Big|_{\,\rm ret}\!V(x_2)\;\frac1{m^2_3-\Box_3}\;\Big|_{\,\rm ret}\!V(x_3)\;\right]_{\,x_2=x_3=x},
    \end{eqnarray}
where we remind that ${V}_\phi(\nabla)$ is a local vertex operator, $\delta V(x)/\delta\phi(y)={V}_\phi(\nabla)\delta(x-y)$, and $\rho(m_1,m_2,m_3)$ is the spectral density of the 3-point vertex (\ref{3point}) \cite{spectral}
    \begin{eqnarray}
    \rho(m_1,m_2,m_3)=
    \frac1{\pi\sqrt{\big[\,(m_1+m_2)^2-m_3^2\,\big]\,
    \big[\,m_3^2-(m_1-m_2)^2\big]}}\,.
    \end{eqnarray}

There exist numerous applications of this covariant perturbation theory and Euclidean version of the Schwinger-Keldysh technique to the particle creation phenomena \cite{Mirzabekianvilkov}, to vacuum backreaction of rapidly moving sources in QED \cite{Vilkovisky}, causality of QED in curved spacetime \cite{Shore} and in the theory of evolving quantum black holes \cite{VilkovBH}.

\subsection{Nonperturbative heat kernel asymptotics and nonlocal effective action}
\hspace{\parindent}Covariant perturbation theory is applicable whenever $d\geq 3$ and the potential $V$ is sufficiently small, so that its effective action explicitly features analyticity at $V=0$. Therefore, its serious disadvantage is that it does not allow one to overstep the limits of perturbation scheme and discover non-analytic structures in the action if they exist.

Nonperturbative technique for the heat kernel is based on the approximation qualitatively different from those of the previous sections. Rather than imposing certain smallness restrictions on the background fields we consider them rather generic, but take into account the both limits of small and large proper time $s\to\infty$ in the heat kernel. This would allow us to capture the effects of local ultraviolet nature and nonlocal infrared effects.

Continuing working in flat spacetime with
$g_{\mu\nu}=\delta_{\mu\nu}$, we substitute the ansatz (\ref{ansatz}) in the heat equation and assume the existence of the following $1/s$-expansion for
$\Omega(s|\,x,y)$ (which follows, in particular, from the perturbation theory for $K(s|\,x,y)$ \cite{CPTII} briefly reviewed above -- there is no nonanalytic terms in $1/s$ like $\ln(1/s)$),
    \begin{equation}
     \Omega(s|\,x,y)=
     \Omega_0(x,y)+\frac1s\,\Omega_1(x,y)
     +O\left(\,\frac1{s^2}\,\right).           \label{3.2}
    \end{equation}
As a result we obtain the series of recurrent equations for $\Omega_n$ starting with \cite{nnea,nneag,TMF}
    \begin{eqnarray}
     &&F(\nabla)\,\Omega_0(x,y)=0,\quad
     F(\nabla)\,\Omega_1(x,y)=
     (x-y)^\mu\nabla_\mu
     \Omega_0(x,y).                     \label{3.4}
    \end{eqnarray}
Interesting peculiarity of this late-time expansion is that the related expansion for the functional {\it trace} of the heat kernel corresponding to (\ref{3.2}) turns out to be
    \begin{equation}
     {{\rm Tr}}K(s)=\frac1{(4\pi s)^{d/2}}\,
     \left\{\,s\,W_0+W_1
     +O\left(\,\frac1s\,\right)\,\right\}.  \label{3.5}
    \end{equation}
This obviously implies that in spite of (\ref{1.11}) $W_n\neq\int dx\:\Omega_n(x,x)$, $n=0,1,...$, because of the unit shift in the power of $s$. This visible mismatch between (\ref{3.2}) and (\ref{3.5}) follows from the fact that the
$1/s$-expansion (\ref{3.2}) is not uniform in $x$ and $y$
arguments of $\,\Omega(s|\,x,y)$. For fixed $s$ the asymptotic
expression $\Omega(s|\,x,x)$ fails to be correct for $|x|\to\infty$, and the heat kernel functional trace (requiring integration up to infinity) cannot be obtained by applying (\ref{1.11}) to (\ref{3.2}). Alternatively, ${\rm Tr}K(s)$ can be recovered from the expansion of $K(s)$ due to the following variational equation
    \begin{equation}
     \frac{\delta \,{{\rm Tr}}K(s)}{\delta V(x)}
     =-sK(s|x,x),                            \label{3.6}
    \end{equation}
which explains, in particular, one extra power of the proper time in (\ref{3.5}) as compared to (\ref{3.2}). This equation generates a sequence of variational equations for $W_n$
    \begin{eqnarray}
     \frac{\delta \,W_n}{\delta V(x)}
     =-\Omega_n(x,x), \qquad n=0,1,...\,.          \label{3.7}
    \end{eqnarray}

The solution of the above equations in the first two orders of late time expansion was obtained in \cite{nnea} in terms of a special function
    \begin{equation}
     \Phi (x)=1+\frac{1}{\Box-V}\,V(x)
     \equiv1+\int dy\:G(x,y)V(y).            \label{3.8}
    \end{equation}
This is a zero mode of the operator $F(\nabla)$ which exists due to unit boundary conditions at infinity, $F(\nabla)\,\Phi(x)=0$, $\Phi (x)\rightarrow 1$, $|x|\rightarrow\infty$. In terms of $\Phi(x)$ the solution of Eqs. (\ref{3.4}) has the form
    \begin{eqnarray}
    &&\Omega_0(x,y)=\Phi(x)\,\Phi(y),     \label{3.10}\\
    &&\Omega_1(x,y)=\frac1{\Box_x-V_x}\,(x-y)^\mu
     \nabla_\mu\Phi(x)\,\Phi(y)\nonumber\\
    &&\qquad\qquad\qquad\qquad
     +\frac1{\Box_x-V_x}\nabla_\mu\Phi(x)\,
     \frac1{\Box_y-V_y}
     \nabla^\mu\Phi(y)
     +(x\leftrightarrow y),                      \label{3.14}
    \end{eqnarray}
which in its turn via Eq.(\ref{3.7}) gives rise to  \cite{nnea}
    \begin{eqnarray}
    &&W_0=-\int dx\:V\,\Phi(x),\quad W_1=\int dx\,\left\{1
     -2\,\nabla_\mu\Phi\,\frac1{\Box-V}\,
     \nabla^\mu\Phi\,\right\}.        \label{3.15}
    \end{eqnarray}

As we saw in Sect.2.2 only for extremely slow and physically
uninteresting falloff with $p<2$ the deviation from homogeneity can be treated by perturbations. For a faster decrease at $|x|\to\infty$ the modified gradient expansion fails.
However, the late time heat kernel asymptotics can give nonlocal and nonperturbative action which captures in
a nontrivial way the edge effects of a transition domain between the regime of finite $|x|$ and the regime of vanishing potential at $|x|\to\infty$. The method consists in taking the two simple functions ${{\rm Tr}}{K}_{<}(s)$ and ${{\rm Tr}}{K}_{>}(s)$
    \begin{eqnarray}
     &&{{\rm Tr}}K_{<}(s)=\frac{1}{(4\pi s)^{2}}
     \int dx\,e^{-sV}, \quad
     {{\rm Tr}}K_{>}(s)=\frac{1}{(4\pi s)^{2}}
     \int dx\,(1-sV\Phi),                    \label{4.2}
    \end{eqnarray}
which coincide with the leading behavior of ${{\rm Tr}}K(s)$ at
$s\rightarrow 0$ and $s\rightarrow \infty$ and using them to
approximate ${{\rm Tr}}K(s)$ respectively at $0\leq s\leq s_{\ast}$ and
$s_{\ast }\leq s<\infty$ for some $s_{\ast }$.
The value of $s_{\ast}$ will be determined from the requirement
that these two functions match at $s_{\ast}$, which will guarantee
the stationarity of $\varGamma$ with respect to the choice of
$s_{\ast}$, $\partial \bar{\varGamma}/\partial s_{\ast}=0$, ${\rm Tr} K_{<}(s_{\ast})={\rm Tr} K_{>}(s_{\ast})$. As shown in \cite{nnea} this piecewise-smooth approximation is efficient at least for two rather wide classes of potentials $V(x)$. They have finite amplitude $V_0$ within their compact support $\cal D$ of size $L$ \cite{nnea},
    \begin{eqnarray}
     &&V(x)=0,\quad |x|\geq L,                 \nonumber\\
     &&V(x)\sim V_0,\quad |x|\leq L,\quad x\in\cal D,           \label{4.6}
    \end{eqnarray}
and have the property that their derivatives are not too high and
uniformly bounded by the quantity of the order of magnitude
$V_0/L$.

For the class of potentials small in units of the inverse
size of their compact support, $V_0 L^2\ll 1$,
the finite part of the action, which is valid up to corrections
proportional to this smallness parameter, reads
    \begin{eqnarray}
    &&\varGamma\simeq \frac{1}{64\pi ^2}
    \int d^4x\,V^2\,\ln \left[\,
    \frac{\int d^4x\,V^2}{\int d^4x\,V\frac{\mu^2}
    {V-\Box}V}\,\right]\,.                \label{4.8}
    \end{eqnarray}
Here we disregard the ultraviolet divergent part of the
action and absorb all finite renormalization type terms $\sim\int d^4x\,V^2$ in the redefinitions of $\mu^2$.

Note, that this renormalization mass parameter $\mu^{2}$ makes the argument of the logarithm dimensionless and plays the same role as for the Coleman-Weinberg potential. However, the original Coleman-Weinberg term for small potentials of the type (\ref{4.6}) gets replaced by the other qualitatively new nonlocal structure. For small potentials spacetime gradients dominate over their magnitude and, therefore, the Coleman-Weinberg term does not survive in this approximation. Still it can be recovered in the formal limit of the constant potential, when the denominator of the logarithm argument tends to $\mu^2\int d^4x\,V$ and the infinite volume factor ($\int d^4x$) gets canceled in the argument of the logarithm.

Another class of potentials, when the piecewise smooth approximation is effective, corresponds to the opposite limit, $V_0 L^2\gg 1$,
that is big potentials in units of the inverse size of their
support $\cal D$. In this case spacetime gradients do not dominate the
amplitude of the potential and the calculation shows
that the effective action contains the Coleman-Weinberg term
modified by the special nonlocal correction \cite{nnea}
    \begin{equation}
    \varGamma\simeq\varGamma_\mathrm{CW}+
    \frac{\left[\,\int\limits_{\cal D}d^4x\,V\Phi\,\right]^2}
    {64\pi^2\int\limits_{\cal D} d^4x}\;.          \label{4.11}
    \end{equation}
Again this algorithm correctly stands the formal limit of a
constant potential, because in this limit the function $\Phi(x)$ given by (\ref{3.8}) formally tends to zero (and the size $L$ grows to infinity).

\subsection{Inclusion of gravity}
\hspace{\parindent} The nonperturbative late-time asymptotics can be nearly straightforwardly generalized to curved spacetime. The flat metric gets replaced by the curved one and the interval in the ansatz (\ref{ansatz}) goes over to the world function -- one half of the geodesic distance squared between the points $x$ and $y$, $\delta_{\mu\nu}
\rightarrow g_{\mu\nu}(x)$, $|x-y|^2/2\rightarrow\sigma(x,y)$.
$\Omega(s|\,x,y)$ instead of (\ref{ansatz})-(\ref{2.3}) has a small-time limit $\Omega(s|\,x,y)\rightarrow g^{-1/2}(x)
     \left(\det{\partial_\mu^x\partial_\nu^y
       \sigma(x,y)}\right)
     g^{1/2}(y)\neq 0$, $s\to 0$,
in terms of the Pauli-Van Vleck-Morette determinant
\cite{PhysRep,DeWitt}. In the assumption of asymptotic flatness, which in cartesian coordinates implies the following falloff metric behavior characteristic of $d$-dimensional Euclidean spacetime
    \begin{eqnarray}
    g_{\mu\nu}(x)\,\Big|_{\,|x|\to\infty}
    =\delta_{\mu\nu}
    +O\left(\frac1{|x|^{d-2}}\right),    \label{5.4}
    \end{eqnarray}
the leading order of late-time expansion for $K(s|\,x,y)$ turns out to be a direct covariantization of the flat-space result.
Almost the same situation holds for the functional trace. Its leading order is given by two terms \cite{nneag,TMF},
    \begin{eqnarray}
     W_0=-\int dx\,&&g^{1/2}\,V\,\Phi
     +\frac16\,\Sigma,                         \label{5.7}\\
    \Sigma=\int dx\,g^{1/2}\,&&\left\{\,R
    -\,R_{\mu\nu}\frac1{\Box} \Big(R^{\mu\nu}
    -\frac12\,g_{\mu\nu}R\Big)
    +\frac12\,R\left(\frac1{\Box}
    R^{\mu\nu}\right)
    \frac1{\Box} R_{\mu\nu}\right.\nonumber\\
       &&-R^{\mu\nu}\left(\frac1{\Box}
    R_{\mu\nu}\right)\frac1{\Box} R
       +\left(\frac1{\Box} R^{\alpha\beta}\right)
    \left(\nabla_\alpha\frac1{\Box} R\right)
    \nabla_\beta\frac1{\Box} R\nonumber\\
    &&-2\,\left(\nabla^\mu\frac1{\Box} R^{\nu\alpha}\right)
    \left(\nabla_\nu\frac1{\Box}
    R_{\mu\alpha}\right)\frac1{\Box} R \nonumber\\
    &&
    -\left.2\,\left(\frac1{\Box} R^{\mu\nu}\right)
    \left(\nabla_\mu\frac1{\Box}
    R^{\alpha\beta}\right)\nabla_\nu\frac1{\Box}
    R_{\alpha\beta}
    +\mathrm{O}[\,R_{\mu\nu}^4\,]\,\right\}.     \label{5.15}
    \end{eqnarray}
One of them, obtained by the variational procedure, is a straightforward covariantization of $W_0$ from Eq.(\ref{3.15}) with the function $\varPhi(x)$ based on the Green's function of the curved space operator $\Box-V$. Another one follows from covariant perturbation theory \cite{CPTII} and, as shown in \cite{TMF}, turns out to be the surface integral over spacetime infinity based on the asymptotically-flat properties of its metric, $g^\infty_{\mu\nu}(x)=\delta_{\mu\nu}
+h_{\mu\nu}(x)\,|_{\,|x|\to\infty}$,  $h_{\mu\nu}(x)\sim 1/|x|^{d-2}$, $\,|x|\to\infty$. This surface integral is linear in perturbations (contributions of higher powers of $h_{\mu\nu}$ to this integral vanish) and involves only a \emph{local} asymptotic behavior of the metric
    \begin{eqnarray}
    \Sigma=\Sigma[\,g_\infty\,]\equiv
    \int\limits_{|x|\to\infty}\!
     d\sigma^\mu\;\delta^{\alpha\beta}
     \Big(\partial_\alpha
     g_{\beta\mu}-
     \partial_\mu g_{\alpha\beta}\Big).      \label{5.8}
    \end{eqnarray}
Here $d\sigma^\mu$ is the surface element on the sphere of radius $|x|\to\infty$, $\,\partial^\mu=\delta^{\mu\nu}\partial_\nu$ and
$h=\delta^{\mu\nu}h_{\mu\nu}$. Thus, the correct expression for $W_0$ is modified by the the surface integral $\Sigma\,[\,g_\infty\,]$, and this integral does not contribute to the metric variational derivative $\delta W_0/\delta g_{\mu\nu}(x)$ at finite $|x|$.

For asymptotically-flat metrics with a power-law falloff at
infinity $h_{\mu\nu}(x)\sim M/|x|^{d-2}$, $\,|x|\to\infty$, the surface integral $\Sigma\,[\,g_\infty]$ forms the contribution to the Einstein action
    \begin{eqnarray}
    S_\mathrm{E}[\,g\,]\equiv -\int dx\,g^{1/2}\,R(g)
      +\Sigma\,[\,g_\infty],     \label{5.17}
    \end{eqnarray}
which guarantees the correctness of the variational procedure
leading to Einstein equations. Covariantly this integral can also be rewritten in the Gibbons-Hawking form
$S_{GH}\,[\,g\,]=\Sigma\,[\,g_\infty]$ -- the double of the
extrinsic curvature trace $K$ on the boundary (with a properly
subtracted infinite contribution of the flat-space background)
\cite{GH}. Thus, this is the surface integral of the \emph{local} function of the boundary metric and its normal derivative. The virtue of the relations (\ref{5.15})-(\ref{5.8}) is that they express this surface integral in the form of the spacetime (bulk) integral of the \emph{nonlocal} functional of the bulk metric. The latter does not explicitly contain auxiliary structures like the vector field normal to the boundary, though these structures are implicitly encoded in boundary conditions for nonlocal operations in the bulk integrand of (\ref{5.15}).

Note also, in passing, that the relation (\ref{5.15}) can be used to rewrite the (Euclidean) Einstein-Hilbert action (\ref{5.17}) as the \emph{nonlocal} curvature expansion which begins with the \emph{quadratic} order in curvature. As will be discussed below, this observation serves as a basis for covariantly consistent nonlocal modifications of Einstein theory \cite{covnonloc} motivated by the cosmological
constant and cosmological acceleration problems \cite{AHDDG}.

\section{Nonlocal cosmology}
\hspace{\parindent}In recent years cosmology became the arena of applications of nonlocal field theory. Major motivation for that was and still is the search for an infrared modification of Einstein general relativity as a model of dark matter and dark energy in the modern Universe \cite{acceleration} and a UV modification as a consistent model of the early quantum Universe. Nonlocal cosmology is descending from the old approach to nonlocal QFT and quantum gravity \cite{Efimov,Tomboulis}, and its latest development embraces its various field-theoretical issues and applications. A very incomplete list of selected works on these issues starts with \cite{SoussaWoodard} and \cite{nonloccosm} and includes the search for structure formation mechanisms \cite{DeffWood,TsamisWoodard,Khoury,Koivisto,Odintsovetal,ParkDodelson}, dynamical screening of the cosmological constant by infrared quantum effects \cite{WoodardIR,AntoniadisMazurMottola}, search for singularity free/bouncing cosmologies \cite{Mazumdaretal}, construction of string inspired nonlocal cosmology \cite{string-inspired}, analysis of renormalizability and unitarity \cite{latest}, etc.

These works treat nonlocal  gravity mostly as an effective theory, nonlocality of effective equations of motion arising from quantizing a fundamental local field theory or string theory. Unfortunately, however, thus far there is no generally recognized mechanism of nonlocal quantum corrections to Einstein equations that could be responsible for a variety of cosmological phenomena. For instance, infrared effects of graviton creation \cite{WoodardIR} that served as a motivation for the nonlocal cosmology of \cite{nonloccosm} stumble upon a serious criticism of \cite{GarrigaTanaka}, string theory implications are also far from forming a reliable mature theory and so on. To the same extent nonlocal terms of effective action discussed in the previous sections also serve merely as a qualitative hint for the type of nonlocality useful for explanation of dark energy or dark matter phenomena. For this reason below we accept somewhat different strategy -- instead of starting with the fundamental theory we will choose a certain structure of a nonlocal model and check its cosmological predictions. This choice will be biased by the necessity to explain dark energy driving the cosmic acceleration \cite{acceleration}.

As is known, dark energy models like \cite{quintessence,f(R),DGP,Deffayet,massive,nonloccosm} suffer from the fine tuning problem associated with the hierarchy of the horizon vs the Planck scale. Modulo certain exceptions \cite{Shaposhnikovetal}, most of them in fact look as a sophisticated way to incorporate the horizon scale (whether it is a graviton mass of massive gravity \cite{massive}, multi-dimensional Planck mass or the DGP scale in brane models \cite{DGP}, etc.). This difficulty can, perhaps, be circumvented by the following line of reasoning \cite{serendipity}. If {\em a concrete fixed scale} incorporated in the model is not satisfactory, then one could look for a model that admits cosmic acceleration with {\em an arbitrary scale}. Then its concrete observable value -- a free parameter of the background solution of equations of motion -- should arise dynamically by the analogue of symmetry breaking to be considered separately. Even this very unassuming approach is full of difficulties, because modified gravity models have additional degrees of freedom which generally lead to ghost instabilities and make the theory inconsistent. This problem is central to numerous attempts to modify Einstein theory, and it will be a major question of this section.

Thus we consider a nonlocal modification of the metric sector of the theory, which is likely to implement this approach. It is based on the realization of the old idea of a scale-dependent gravitational coupling -- nonlocal Newton ``constant" \cite{AHDDG,covnonloc,HamberWilliams} -- and  amounts to the construction of the class of ghost free models compatible with the GR limit and generating the de Sitter (dS) or anti-de Sitter (AdS) background with an {\em arbitrary value} of its effective cosmological constant $\varLambda$ \cite{serendipity}.

\subsection{Scale dependent coupling -- nonlocal gravitational ``constant"}
\hspace{\parindent}The concept of the effective scale dependent gravitational constant was introduced in \cite{AHDDG} as an implementation of the idea that the effective cosmological constant in modern cosmology is very small not because the vacuum energy of quantum fields is so small, but rather because it gravitates too little. This degravitation is possible if the effective gravitational coupling ``constant" depends on the momentum and becomes small for fields nearly homogeneous at the horizon scale. Naive replacement of the Newton constant by a nonlocal operator suggested in \cite{AHDDG} violates diffeomorphism invariance, but this procedure can be done covariantly due to the following observation \cite{covnonloc}.

The Einstein action in the vicinity of a flat-space background can be rewritten in the form starting with the nonlocal term bilinear in Ricci tensor and Einstein tensor, $G_{\mu\nu}=R_{\mu\nu}-\frac12 g_{\mu\nu}R$,
    \begin{equation}
    S_E=
    \frac{M_P^2}2\int dx\,g^{1/2}\,\left\{\,
    -R^{\mu\nu}\frac1{\Box}\,G_{\mu\nu}\,
    +{\rm O}\,[R_{\mu\nu}^3]\,\right\},       \label{flatE}
    \end{equation}
where $1/\Box$ is the Green's function of the covariant d'Alembertian acting on a symmetric tensor of second rank. This expression is nothing but a generally covariant version of the quadratic part of the Einstein action in metric perturbations $h_{\mu\nu}$ on a flat space background. When rewritten in terms of the Ricchi tensor $R_{\mu\nu}\sim \nabla\nabla h+O[h^2]$ this expression becomes nonlocal but preserves diffeomorphism invariance to all orders of its curvature expansion. This expression for the Einstein action follows from the subtraction of the linear in scalar curvature term by the surface Gibbons-Hawking integral over asymptotically-flat infinity
    \begin{eqnarray}
    &&S_E=-\frac{M_P^2}2
    \int dx\,g^{1/2}\,R(\,g\,)+\frac{M_P^2}2
    \int_\infty d\sigma^\mu\,
    \big(\partial^\nu
    h_{\mu\nu}-\partial_\mu h).        \label{GHsubtraction}
    \end{eqnarray}
As discussed in Sect. 2.4, this surface term is a topological invariant depending only on the asymptotic behavior $g^\infty_{\mu\nu}=\delta_{\mu\nu}+
h_{\mu\nu}(x)\,|_{\,|x|\to\infty}$.  According to Eqs.(\ref{5.15}) and (\ref{5.8}) it can be converted into the form of the volume integral and covariantly expanded in powers of the curvature. This expansion starts with \cite{TMF}
    \begin{eqnarray}
    &&\int_\infty\! d\sigma^\mu\,
    \big(\partial^\nu
    h_{\mu\nu}-\partial_\mu h\Big)=
    \int dx\,g^{1/2}\left\{R
    -R^{\mu\nu}\frac1{\Box}\,G_{\mu\nu}
    +{\rm O}\,[\,R_{\mu\nu}^3\,]\right\},     \label{GH}
    \end{eqnarray}
so that the Ricci scalar term gets canceled in (\ref{GHsubtraction}) and we come to (\ref{flatE}).

With this new representation of the Einstein action, the idea of a nonlocal scale dependent Planck mass \cite{AHDDG} consists in the replacement of $M_P^2$ in (\ref{flatE}) by a nonlocal operator -- a function $M^2(\Box)$ of $\Box$ sandwiched between the Ricci and Einstein tensors,
    \begin{equation}
    M_P^2 R^{\mu\nu}\frac1\Box\,G_{\mu\nu}\to\,
    R^{\mu\nu}\frac{M^2(\Box)}{\Box}\,G_{\mu\nu}.
    \end{equation}
It would realize this idea at least within the lowest order of the covariant curvature expansion and would lead to degravitation in the infrared limit if one assumes a weakening gravitational interaction of the homogeneous sources, $M^2(\Box)\equiv 1/8\pi G(\Box)\to\infty$ at $\Box\to 0$.\footnote{Fading gravity behavior of \cite{Khoury} implies a complementary asymptotic behavior $M^2(\Box)\to 0$ at $\Box\to\infty$ and corresponds to singularity free gravity in UV limit of \cite{Mazumdaretal}.} This modification put forward in \cite{AHDDG,covnonloc} did not, however, find interesting applications because it has left unanswered a critical question -- is this construction free of ghost instabilities for any nontrivial choice of $M^2(\Box)$?

\subsection{Problem of ghosts}
\hspace{\parindent}The search for a consistent $M^2(\Box)$ should be supervised by the correspondence principle -- nonlocal terms of the action should form a correction to the Einstein Lagrangian arising via the replacement $R\to R+R^{\mu\nu}F(\Box)G_{\mu\nu}$. The nonlocal form factor of this correction $F(\Box)$ should be small in the GR domain, but it must considerably modify dynamics at the DE scale. Motivated by customary spectral representations for nonlocal quantities like
    \begin{eqnarray}
    F(\Box)=\int dm^2\,\frac{\alpha(m^2)}{m^2-\Box}
    \end{eqnarray}
we might try the following ansatz, $F(\Box)=\alpha/(m^2-\Box)$, corresponding to the spectral density $\alpha(m^2)$ sharply peaked around some $m^2$ (cf. a similar discussion in \cite{DvaliHofmannKhoury}). For $m^2\neq 0$ this immediately leads to a serious difficulty. Schematically the inverse propagator of the theory becomes
    \begin{eqnarray}
    -\Box+\alpha\frac{\Box^2}{m^2-\Box},
    \end{eqnarray}
where the squared d'Alembertian $\Box^2$ follows from four derivatives contained in the term bilinear in curvatures.
Then its physical modes are given by the two roots of this expression -- the solutions of the corresponding quadratic equation $\Box=m_\pm^2$. In addition to the massless graviton with $m_-^2=0$ massive modes with $m_+^2=O(m^2)$ appear and contribute a set of ghosts which cannot be eradicated by gauge transformations (for the latter have to be expended on cancelation of ghosts in the massless sector -- longitudinal and trace components of $h_{\mu\nu}$ subject to $\Box h_{\mu\nu}=0$.).

Therefore, only the case of $m^2=0$ remains, and as a first step to the nonlocal gravity we will consider the action
    \begin{eqnarray}
    S=
    \frac{M^2}2\int dx\,g^{1/2}\,\left\{\,-R+
    \alpha\,R^{\mu\nu}
    \frac1\Box\,G_{\mu\nu}\,\right\}            \label{action0}
    \end{eqnarray}
(for brevity we omit the surface integral that should accompany the Einstein Ricci scalar term). On the flat-space background this theory differs little from GR provided the dimensionless parameter $\alpha$ is small, $|\alpha|\ll 1$. Upper bound on $|\alpha|$ should follow from post-Newtonian corrections in this model. The additional effect of $\alpha$ is a small renormalization of the effective Planck mass -- the linearized limit of the theory allows one to relate the constant $M$ to $M_P$ by
    \begin{eqnarray}
    M^2=\frac{M^2_P}{1-\alpha},  \label{M_Prenorm}
    \end{eqnarray}

As we will see later, application of the model (\ref{action0}) in cosmological setup fails due to inconsistent treatment of boundary conditions, and it is instructive to see their importance. Like in papers on $f(R/\Box)$-gravity (see \cite{Odintsovetal} and references therein) stemming from \cite{nonloccosm}, but in contrast to \cite{nonloccosm} disregarding consistent treatment of boundary conditions, one can localize the nonlocal part of (\ref{action0})  with the aid of an auxiliary tensor field $\varphi^{\mu\nu}$. Then, the theory is equivalently described by the action
    \begin{eqnarray}
    &&S[\,g,\varphi\,]=
    \frac{M^2}2\int dx\,g^{1/2}\,\Big\{-R
    -2\alpha\,\varphi^{\mu\nu}R_{\mu\nu}
    -\alpha\,\Big(\varphi^{\mu\nu}
    -\frac12\,g^{\mu\nu}\varphi\Big)\,
    \Box\,\varphi_{\mu\nu}\Big\}            \label{varphiaction}
    \end{eqnarray}
generating for $\varphi^{\mu\nu}$ the equation of motion  $\Box\varphi^{\mu\nu}=-G^{\mu\nu}$. Referring to Sect.2.2 let us interpret the expression (\ref{action0}) as the Euclidean spacetime action with zero boundary conditions for $1/\Box$ at infinity. Then the auxiliary tensor field should satisfy the same Dirichlet boundary conditions $\varphi^{\mu\nu}|_{\,\infty}=0$, and this is critically important for stability of the theory. Indeed, the field $\varphi^{\mu\nu}$ formally contains ghosts, but they do not indicate physical instability because they never exist as a free fields in the external lines of Feynman graphs. In the Lorentzian context of (\ref{EuclidLorentz}) this means that $\varphi^{\mu\nu}$ is given by a retarded solution, $\varphi^{\mu\nu}=-(1/\Box)_{\rm ret}G^{\mu\nu}$, and does not include free waves coming from past infinity.

Artificial nature of these ghosts is analogous to the case of the simplest ghost-free action that can be formally rendered nonlocal
    \begin{equation}
    S[\,\varphi\,]\equiv
    -\int dx\,\varphi \Box\varphi=
    -\int dx\,(\Box\varphi)
    \frac1{\Box}(\Box\varphi)    \nonumber
    \end{equation}
and further localized in terms of the auxiliary field $\psi$ with the action $S[\,\varphi,\psi\,]=
    \int dx\,\left(2\psi\,\Box\varphi
    +\psi\,\Box\psi\right)$.
This action is equivalent to the original one when $\psi$ is integrated out with the boundary conditions $(\psi+\varphi)\,|_{\,\infty}=0$.
After diagonalization this action features the ghost field $g\equiv\psi+\varphi$, $S[\,\varphi,\psi\,]=
    \int dx\, \left(g\,\Box g
    -\varphi\,\Box\varphi\right)$.
This ghost is however harmless because under the boundary conditions of the above type it identically vanishes in view of its equation of motion $\Box g=0$. In the presence of interaction, a nonvanishing $g$ exists in the intermediate states, but never arises in the asymptotic states, or external lines of Feynman graphs.

Main lesson to be drawn from the above example is that the actual particle content of the theory should be determined in terms of the original set of fields, whereas nonlocal reparameterizations can lead to artificial ghost modes which are actually eliminated by correct boundary conditions. In our case this is the original formulation (\ref{action0}) in terms of the metric field $g_{\mu\nu}$. It indeed turns out to be ghost-free on the flat-space background, because the quadratic part of the action coincides with the Einstein one.

A formal application of (\ref{action0}) in the FRW setup disregards nontrivial boundary conditions in cosmology. To see this, note that initial conditions for DE data would generally contradict zero boundary conditions for the auxiliary tensor field $\varphi^{\mu\nu}$, not to mention that the cosmological FRW setup does not in principle match with the asymptotically-flat framework of the action (\ref{action0}). Therefore we have to extrapolate the definition (\ref{action0}) to nontrivial backgrounds including, first of all, the de Sitter spacetime and change our technique -- instead of localization method with an auxiliary tensor field work directly in the original metric representation. Then immediately a serious difficulty arises. Ricci curvature for the (A)dS background is covariantly constant, and the nonlocal part of (\ref{action0}) turns out to be infrared divergent, $(1/\Box)G_{\mu\nu}=\infty$. This means that the action (\ref{action0}) should be modified to regulate this type of divergences which will be done after we consider the causality problem.

\subsection{Problem of causality}
\hspace{\parindent}Now we have to address the treatment of nonlocality in (\ref{action0}) and (\ref{action}). Handling the theories with a nonlocal action is a sophisticated and very often an open issue, because their nonlocal variational equations of motion demand special care in setting their boundary value problem \cite{nonlocbc}. Contrary to local field theories subject to a standard Cauchy  problem setup and canonical commutation relations, nonlocal theories can have very ambiguous rules which are critical for physical predictions. In particular, the action (\ref{action0}) above requires specification of boundary conditions for the nonlocal Green's function $1/\Box$ which will necessarily violate causality in variational equations of motion for this action. Indeed, the action (\ref{action0}) effectively symmetrizes the kernel of the Green's function $G(x,y)$ of $1/\Box$, so that nonlocal terms in variational equations of motion
    \begin{equation}
    \frac{\delta S}{\delta g_{\mu\nu}(x)}\propto
    \nabla\nabla\int dy\,\big[\,G(x,y)+G(y,x)\,\Big]\,{\cal R}(y)+...
    \end{equation}
(${\cal R}(y)$ denoting a collection of curvatures) never have retarded nature even when $G(x,y)$ is the retarded propagator or satisfies any other type of boundary conditions \cite{DeffWood,serendipity}. Therefore, these equations break causality because the behavior of the field at the point $x$ is not independent of the field values at the points $y$ belonging to the future light cone of $x$, $y^0>x^0$.

To avoid these ambiguities we assume that our nonlocal action is not fundamental but rather represents the quantum effective action -- the generating functional of one-particle irreducible diagrams -- whose argument is the mean quantum field. As discussed above in Sect.2.2 this functional is uniquely determined by the choice of the mean field (either IN-OUT or expectation value IN-IN) and the relevant boundary conditions are uniquely fixed  by the choice of the initial (and/or final) quantum state. In what follows we will be interested in the IN-IN expectation value (\ref{meanfield}) of the metric field $\phi(x)=g_{\mu\nu}(x)\equiv \langle\,{\rm in}\,|\,\hat g_{\mu\nu}(x)\,|\,{\rm in}\,\rangle$ which should satisfy the causality condition -- retarded dependence on its classical and quantum sources $J(x)$ (including the self-interaction ones),
    \begin{eqnarray}
    \frac{\delta\langle\,{\rm in}\,|\,\hat g_{\mu\nu}(x)\,|\,{\rm in}\,\rangle}{\delta J(y)}=0,\quad x^0<y^0.
    \end{eqnarray}
This means that the radiation current (\ref{1loopradcur}) in effective equations for $g_{\mu\nu}(x)$, $J_{IN}^{\mu\nu}(x)[\,g_{\alpha\beta}(y)\,]$ should also have a retardation type functional dependence on $g_{\alpha\beta}(y)$, $\delta J_{IN}^{\mu\nu}(x)/\delta g_{\alpha\beta}(y)=0$ for $y^0>x^0$.

For the calculation of $J_{IN}^{\mu\nu}(x)$ one can use the Euclidean version of the Schwinger-Keldysh technique discussed in Sect.2.2. Let us remind that it starts with the calculation of the Euclidean effective action $\varGamma_{\rm Euclid}[\,g_{\mu\nu}\,]$ and its variational derivative. In the Euclidean signature spacetime nonlocal quantities, relevant Green's functions and their variations are generally uniquely determined by their trivial (zero) boundary conditions at infinity, so that this variational derivative is unambiguous in Euclidean theory. Then a formal transition to the Lorentzian signature with imposed {\em retarded} boundary conditions on the resulting nonlocal operators gives the final form of effective equations
    \begin{eqnarray}
    \left.\frac{\delta\varGamma_{\rm Euclid}}{\delta g_{\mu\nu}}\right|_{\;++++\,\;
    \to\;-+++}^{\;\rm retarded}=0.   \label{EuclidLorentz}
    \end{eqnarray}
They are causal ($g_{\mu\nu}(x)$ depending only on the field behavior in the past of the point $x$) and satisfy all local gauge and diffeomorphism symmetries encoded in the original $\varGamma_{\rm Euclid}[\,g_{\mu\nu}\,]$.

This retardation version of Wick rotation algorithm was proven in \cite{CPTI} only for asymptotically-flat spacetime with an initial state in the form of the Poincare-invariant vacuum and only in the one-loop approximation. The boundary conditions on the Euclidean side of this algorithm are obviously the Dirichlet ones at spacetime infinity. However, recent results of \cite{HiguchiMarolfMorrison,Tanaka} apparently extend this algorithm to the perturbation theory in the open chart of the de Sitter spacetime for the de Sitter invariant vacuum state. Remarkably, the situation with boundary conditions becomes nontrivial -- despite open spatially flat chart of the physical de Sitter spacetime, its Euclidean counterpart is a closed compact sphere $S^4$ which imposes as Euclidean boundary conditions nothing but requirements of regularity. This implies that when calculating the Euclidean effective action within perturbation theory on $S^4$-background one can freely integrate by parts without generating surface terms -- the analogue of the property guaranteed in asymptotically flat case by Dirichlet conditions at infinity.

\begin{figure}[h]
\centerline{\epsfxsize 12cm \epsfbox{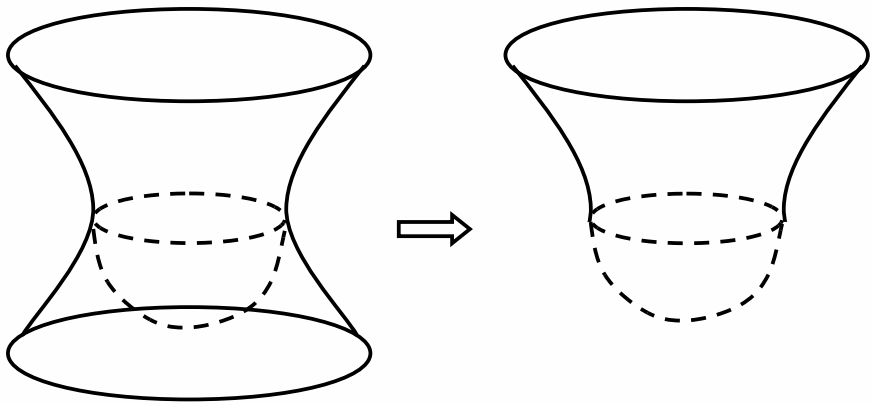}} \caption{\small
Euclidean de Sitter hemisphere denoted by dashed lines is used as a tool for constructing the Euclidean de Sitter invariant vacuum by the path integral over regular fields on the resulting compact spacetime.
 \label{Fig.1}}
\end{figure}

At the heuristical level the justification for this extension follows from Fig.\ref{Fig.1} depicting the compact Euclidean spacetime used as a tool for constructing the Euclidean vacuum within a well-known no-boundary prescription \cite{noboundary}. Attaching a Euclidean space hemisphere to the Lorentzian de Sitter spacetime makes it {\em compact} instead of the original asymptotic de Sitter infinity. Thus it simulates by the path integral over regular field configurations on this spacetime the effect of the Euclidean de Sitter invariant vacuum. The role of spacetime {\em compactness} is very important here because it allows one to disregard possible surface terms originating from integrations by parts or using cyclic permutations under the functional traces in the Feynman diagrammatic technique for the effective action.

In what follows this property will be very important. In particular, the Green's function will be uniquely defined by the condition of regularity on such a compact spacetime without a boundary. This information is sufficient to specify the Green's function of the operator $F=\Box+P$, for which we require the following symmetric variational law (with respect to local metric variations of $\Box$ and the potential term $P$)
    \begin{eqnarray}
    \delta\frac1{\Box+P}=
    -\frac1{\Box+ P}\,\delta\big(\Box+ P\big)
    \frac1{\Box+P},                   \label{symvar}
    \end{eqnarray}
characteristic of the Euclidean signature d'Alembertian (Laplacian) defined on the space of regular fields on a compact spacetime without a boundary. The only important restriction here is the requirement of invertibility of this operator -- absence of zero modes of $F=\Box+P$ (which for example is guaranteed for massive fields with $P=-m^2$ making $F$ negative definite on compact manifolds without a boundary).

A similar method of deriving covariant and causal effective equations of motion from the nonlocal action was conjectured and very reservedly called the ``integration by parts trick" in \cite{nonloccosm,DeffWood}. The justification for it was that it automatically guarantees causality and diffeomorphism Noether identities (conservation of stress tensor) for {\em any} covariant nonlocal action functional, the ``trick" status of this procedure being that the variational procedure and the retardation rule both have a formal ad hoc nature. It was emphasized that this procedure requires justification from the Schwinger-Keldysh technique, and the Euclidean version of this technique of \cite{CPTI,HiguchiMarolfMorrison,Tanaka} seems to put this conjecture on a firm ground.\footnote{Contrary to other papers on $f(\Box^{-1}R)$ approach to nonlocal cosmology, like \cite{ParkDodelson} and the others, which assume the possibility of causal covariant equations of motion not necessarily derivable by the variational procedure. I am grateful to S. Deser and R. Woodard for the discussion of this point.}

\section{Nonlocal gravity as a source of dark energy}
\hspace{\parindent}Motivated by the ghost free nonlocal theory (\ref{action0}) on a flat space background we now go over to the model which might realize a new approach to the dark energy problem avoiding the fine tuning. This approach suggests the theory in which the de Sitter or anti-de Sitter evolution can occur at any value of the effective cosmological constant $\Lambda$ -- the antithesis to the dark energy scale encoded in the action of the model and fine tuned to the observational data.

As noted in the end of Sect.4.2 and the preceding section, to incorporate the (A)dS background the action (\ref{action0}) should be regulated to get rid of the zero mode of $\Box$. This is done by adding to the covariant d'Alembertian the matrix-valued potential term built of a generic combination of tensor structures linear in the curvature \cite{serendipity}. This brings us to the six-parameter family of nonlocal action functionals
    \begin{eqnarray}
    &&S=\frac{M^2}2\int dx\,g^{1/2}\,\left\{-R+
    \alpha\,R^{\mu\nu}
    \frac1{\Box+\hat P}\,G_{\mu\nu}
    \right\},\;\;\;\;                        \label{action}\\
    &&\hat P\equiv P_{\alpha\beta}^{\;\;\;\mu\nu}
    =a R_{(\alpha\;\;\beta)}^{\;\;\,(\mu\;\;\,\nu)}
    +b \big(g_{\alpha\beta}R^{\mu\nu}
    +g^{\mu\nu}R_{\alpha\beta}\big)
    +c R^{(\mu}_{(\alpha}\delta^{\nu)}_{\beta)}
    +d R\,g_{\alpha\beta}g^{\mu\nu}
    +e R \delta^{\mu\nu}_{\alpha\beta},     \label{potential}
    \end{eqnarray}
where the hat denotes matrices acting on symmetric tensors, and we use the condensed notation for the Green's function of the covariant operator
    \begin{eqnarray}
    \Box+\hat P\equiv\Box\,\delta_{\alpha\beta}^{\;\;\;\mu\nu}
    +P_{\alpha\beta}^{\;\;\;\mu\nu},
    \quad
    \Box=g^{\lambda\sigma}
    \nabla_\lambda\nabla_\sigma,  \label{operator}
    \end{eqnarray}
acting on any symmetric tensor field $\Phi_{\mu\nu}$ as
    \begin{eqnarray}
    &&\frac1{\Box+\hat P}\,\Phi_{\mu\nu}(x)\equiv
    \Big[\,\frac1{\Box+\hat P}\,\Big]_{\mu\nu}^{\alpha\beta}\Phi_{\alpha\beta}(x)
    =\int dy\,G_{\mu\nu}^{\alpha\beta}(x,y)\,
    \Phi_{\alpha\beta}(y)
    \end{eqnarray}
with $G_{\mu\nu}^{\alpha\beta}(x,y)$ -- the two-point kernel of this Green's function. These six parameters are restricted by the requirement of a stable (A)dS solution existing in this theory. These restrictions read \cite{serendipity}
    \begin{eqnarray}
    &&\alpha=-A-4B,    \label{relation}\\
    &&C=\frac23,        \label{Crelation}\\
    &&M_{\rm eff}^2
    =\frac{A^2-\alpha^2}{\alpha}\,M^2>0. \label{effectivemass}
    \end{eqnarray}
where the new quantities $A$, $B$ and $C$ equal in terms of original parameters
    \begin{eqnarray}
    &&A=a+4\,b+c,\quad
    B=b+4\,d+e, \quad C=\frac{a}3-c-4e,            \label{C}
    \end{eqnarray}
and $M_{\rm eff}$ is the effective Planck mass which determines the cutoff scale of perturbation theory in the (A)dS phase and the strength of the gravitational interaction of matter sources.

The condition (\ref{relation}) guarantees the existence of the (A)dS solution, while Eqs.(\ref{Crelation})-(\ref{effectivemass}) are responsible for its stability.  The calculation of the gauge fixed quadratic part of the action on the (A)dS background shows that longitudinal and trace modes which formally have a ghost nature are unphysical and can be eliminated by residual gauge transformations preserving the gauge \cite{serendipity}. This well-known mechanism leaves only two transverse-traceless physical modes propagating on the (A)dS background, similar to GR theory. Finally, as was shown in \cite{Solodukhin,newform} the additional condition,
    \begin{eqnarray}
    a=2,                                 \label{arelation} \end{eqnarray}
allows one to extend the ghost stability arguments to generic Einstein backgrounds $R_{\mu\nu}=\Lambda g_{\mu\nu}$ with a nonvanishing Weyl tensor but with a vanishing traceless part of the Ricci tensor
    \begin{eqnarray}
    E_{\mu\nu}\equiv
    R_{\mu\nu}-\frac14\,g_{\mu\nu}R=0.  \label{Einsteinspace}
    \end{eqnarray}

This model formally falls into the category of nonlocal theories descending from the old approach to nonlocal QFT and quantum gravity \cite{Efimov} and its latest development \cite{nonloccosm} motivated by cosmological implications \cite{DeffWood,TsamisWoodard,Odintsovetal,ParkDodelson} and the requirements of renormalizability and unitarity \cite{latest}. However, in contrast to the functional ambiguity in the choice of action, characteristic of these works, here we have only a parametric freedom. Moreover, majority of proposals for a nonlocal gravity theory operate only with the scalar curvature, whereas here we have all curvature components involved.\footnote{Which is justified for a reason similar to the fact that only $R^2+R_{\mu\nu}^2$ theory is UV renormalizable, while pure $R^2$ is not.} The action (\ref{action}) has one dimensional parameter $M$ and six interrelated dimensionless parameters $\alpha$, $a$, $b$, $c$, $d$ and $e$, the first one $\alpha$ determining the overall magnitude of the nonlocal correction to the Einstein term. For a small value of $|\alpha|\ll 1$ and the value of $M$ related to the Planck mass $M_P$ by Eq.(\ref{M_Prenorm}) the theory (\ref{action}) seems to have a GR limit on a {\em flat-space background}, and also in the IR regime has the (A)dS phase at some scale $\varLambda$.\footnote{In fact, as we will see below, interpolation between the GR and (A)dS phases is very subtle and can be achieved only in the conformal frame related to the original metric by a special nonlocal conformal factor \cite{newform}.}

Important property of the action (\ref{action}) is that it homogeneously transforms under global metric dilatations
    \begin{eqnarray}
    S[\,\lambda\,g_{\mu\nu}]
    =\lambda\,S[\,g_{\mu\nu}],     \label{scaling}
    \end{eqnarray}
and this dilatation covariance is in fact the source of the ``indifference" of the theory in the choice of the scale $\varLambda$ in its Einstein space solution. It is clear than that the mechanism of selecting a concrete value of this scale is the breakdown of this transformation law, analogous to conventional spontaneous symmetry breakdown (cf. \cite{Shaposhnikovetal}).

All the above conclusions (\ref{Crelation})-(\ref{C}) regarding the stability of the (A)dS phase of the theory have been reached in \cite{serendipity} by very complicated calculations. However, they can be essentially simplified by noting that the Euclidean action (\ref{action}) with a critical value (\ref{relation}) of $\alpha$ has on a compact manifold without boundary another representation
    \begin{eqnarray}
    &&S=-\frac{M^2_{\rm eff}}2
    \int dx\,g^{1/2}\,E^{\mu\nu}
    \frac1{\Box+\hat P}\,E_{\mu\nu}.  \label{newrep}
    \end{eqnarray}
The advantage of this representation is obvious -- quadratic in $E_{\mu\nu}$ form of (\ref{newrep}) directly indicates the existence of Einstein space solutions satisfying (\ref{Einsteinspace}) and also very easily gives the inverse propagator of the theory on their background. Single-pole nature of the propagator with a positive residue yields the ghost-free criteria (\ref{Crelation})-(\ref{effectivemass}) and (\ref{arelation}). Below we briefly repeat these derivations based on compact and closed nature of the Euclidean spacetime.

\subsection{Compactness of spacetime and stability of Einstein space background}
\hspace{\parindent}The representation (\ref{newrep}) of the action (\ref{action}) follows from the local relation for the operator $\Box+\hat P$ which is valid for an arbitrary scalar function $\varPhi$,
    \begin{eqnarray}
    (\Box+\hat P)\,g_{\mu\nu}\varPhi=g_{\mu\nu}\left(\Box
    -\frac\alpha4\,R\right)\varPhi
    +A\,E_{\mu\nu}\varPhi.         \label{equation0}
    \end{eqnarray}
Putting $\varPhi=1$ and acting by the Green's function $(\Box+\hat P)^{-1}$ on this relation we get the nonlocal identity
    \begin{eqnarray}
    \frac1{\Box+\hat P}\,g_{\mu\nu}\frac{R}4
    =-\frac1\alpha\,g_{\mu\nu}
    +\frac{A}\alpha\frac1{\Box+\hat P}\,
    E_{\mu\nu},                       \label{equation3}
    \end{eqnarray}
because $(\Box+\hat P)^{-1}(\overrightarrow\Box+\hat P)=(\Box+\hat P)^{-1}(\overleftarrow\Box+\hat P)=1$ in view of absence of surface terms. Application of this identity in (\ref{action}) gives (\ref{newrep}).

This immediately allows one to prove the existence of a generic Einstein space solutions (including the maximally symmetric ones derived in \cite{serendipity}) and the absence of ghost modes on top of them. Since (\ref{newrep}) is quadratic in $E_{\mu\nu}$ its first order derivative is at least linear in $E_{\mu\nu}$ with some complicated nonlocal operator coefficient,
    \begin{eqnarray}
    &&\frac{\delta S}{\delta g_{\mu\nu}}=\frac{M^2_{\rm eff}}2
    \,g^{1/2}\,
    \Omega^{\mu\nu}_{\;\;\;\;\alpha\beta}(\nabla)\,
    \frac1{\Box+\hat P}\,
    E^{\alpha\beta},                         \label{eom}\\
    &&\Omega^{\mu\nu}_{\;\;\;\;\alpha\beta}(\nabla)
    =\Box\,\delta^{\mu\nu}_{\alpha\beta}
    +g^{\mu\nu}\nabla_\alpha\nabla_\beta
    -2\nabla_{(\alpha}\nabla^{(\mu}\delta^{\nu)}_{\beta)}
    +\frac12\,R\,
    \delta^{\mu\nu}_{\alpha\beta}+O[\,E\,],   \label{Omega}
    \end{eqnarray}
where $O[\,E\,]$ denotes terms vanishing in the limit $E_{\mu\nu}\to 0$. This guarantees the existence of vacuum solutions with $E_{\mu\nu}=0$. Perturbative stability of these solution follows from the quadratic part of the action, which is easily calculable now.

In view of the quadratic nature of (\ref{newrep}), the quadratic part of the action on the Einstein space background requires variation of only two explicit $E_{\mu\nu}$-factors. For the metric variations $\delta g_{\mu\nu}\equiv h_{\mu\nu}$ satisfying the DeWitt gauge
    \begin{eqnarray}
    \chi^\mu\equiv\nabla_\nu
    h^{\mu\nu}-\frac12\,\nabla^\mu h=0,  \label{DWgauge}
    \end{eqnarray}
the variation of $E_{\mu\nu}$ reads $\delta E_{\mu\nu}\big|_{\;E_{\alpha\beta}=0}=-\frac12\hat D\,
    \bar h_{\mu\nu}$, where the operator $\hat D$
    \begin{eqnarray}
    \hat D\equiv \Box+2\hat W
    -\frac16\,R\,\hat1,       \label{D}
    \end{eqnarray}
acts on a traceless part of $h_{\mu\nu}$,
    \begin{eqnarray}
    &&\bar h_{\mu\nu}\equiv
    \hat\varPi h_{\mu\nu}=h_{\mu\nu}
    -\frac14\,g_{\mu\nu}h,
    \end{eqnarray}
the hat labels matrices acting on symmetric tensors, $\hat\varPi\equiv
    \varPi_{\mu\nu}^{\;\;\alpha\beta}
    =\delta_{\mu\nu}^{\alpha\beta}
    -\frac14\,g_{\mu\nu}g^{\alpha\beta}$,
    $\hat W h_{\mu\nu}\equiv W_{\mu\;\,\nu}^{\;\,\alpha\;\,\beta}
    h_{\alpha\beta},$
and $W_{\mu\;\nu}^{\;\alpha\;\beta}$ denotes the Weyl tensor.
The operator $\hat D$ commutes with the projector $\hat\varPi$, $[\hat\varPi,\hat D]=0$, because of the traceless nature of the Weyl tensor, $\hat\varPi\hat W=\hat W\hat\varPi=\hat W$, so that the variation of the traceless $E_{\mu\nu}$ is also traceless as it should.

In matrix notations the operator $\Box+\hat P$ on the Einstein background reads
    \begin{eqnarray}
    \big(\Box+\hat P\big)\Big|_{\;E_{\mu\nu}=0}=\Box+a\,\hat W-\frac{C}4\,R\hat\varPi
    -\frac\alpha4\,R\,(\hat1-\hat\varPi).
    \end{eqnarray}
Therefore, in view of the property $[\hat\varPi,\hat D]=0$ and the obvious relation $\hat\varPi\,(\Box+\hat P)^{-1}\hat\varPi=\hat\varPi\,(\Box+a\,\hat W-\frac{C}4\,R\,\hat1)^{-1}\hat\varPi$
we finally have the quadratic part of the action in terms of the traceless part $\bar h_{\mu\nu}$ of the metric perturbations $h_{\mu\nu}$ satisfying the DeWitt gauge \cite{Solodukhin,newform}
    \begin{eqnarray}
    S_{(2)}\Big|_{\;E_{\mu\nu}=0}
    =-\frac{M^2_{\rm eff}}2
    \int d^4x\,g^{1/2}\big(\hat D\bar h^{\mu\nu}\big)
    \,\frac1{\Box+a\,\hat W
    -\frac{C}4\,R\,\hat1}\,
    \big(\hat D\bar h_{\mu\nu}\big).  \label{S_2}
    \end{eqnarray}

For generic values of the parameters $a$ and $C$ the propagator of the theory features double poles corresponding to the zero modes of the operator $\hat D$. This is a nonlocal generalization of the situation characteristic of the critical gravity theories with a local action containing higher-order derivatives \cite{critical}. However, flexibility in the values of $a$ and $C$ allows us to avoid perturbative instability of the Einstein space background. This is achieved by demanding equality of the operator $\hat D$ and the operator in the denominator of (\ref{S_2}) along with the positivity of $M^2_{\rm eff}$. This yields the value $C=2/3$ derived in \cite{serendipity} by very extensive calculations and in addition leads to a unique value $a=2$, which allows one to extend stability arguments to generic Einstein space backgrounds \cite{Solodukhin} with a nonvanishing Weyl tensor\footnote{Basic example of a physically nontrivial Einstein space is the Schwarzchild-de Sitter background. A priori it can also generate surface terms in (\ref{equation2}), because its metric is not smooth simultaneously at the black hole and cosmological horizons and has a conical singularity \cite{GibHawkPage}. We show in Appendix, however, that for any type of regular boundary conditions at this singularity the relevant surface term vanishes and leaves Eq.(\ref{equation1}) intact. A similar issue remains open in the case of the Schwarzchild-AdS background for which the operator $\hat D$ with $R<0$ is not guaranteed to be free of zero modes \cite{Solodukhin}. We are grateful to S. Solodukhin for a discussion of this point.}. Then the quadratic form (\ref{S_2}) becomes local and guarantees the existence of the propagator with a single positive-residue pole,
    \begin{eqnarray}
    S_{(2)}[\,\bar h\,]\big|_{\;E_{\mu\nu}=0}
    =-\frac{M^2_{\rm eff}}2
    \int d^4x\,g^{1/2}\,\bar h^{\mu\nu}
    \hat D\,\bar h_{\mu\nu}.                   \label{S_20}
    \end{eqnarray}
Positivity of $M^2_{\rm eff}$ selects two admissible intervals for the parameter $A$ in the case of a positive $\alpha$,$0<\alpha<|A|$, and the compact range of this parameter for a negative $\alpha$, $\alpha<A<-\alpha$.

\subsection{Propagating physical modes and retarded gravitational potentials}
\hspace{\parindent}In cosmology it is the de Sitter background with $\varLambda>0$ which is of interest. It turns out that as free propagating modes it carries the usual GR graviton with two polarizations. To prove it we apply the Euclidean version of the Schwinger-Keldysh technique to (\ref{S_20}). First we rewrite the traceless part of the metric perturbation $\bar h_{\mu\nu}$ in the DeWitt gauge (\ref{DWgauge})), $\chi_\mu(\bar h)=0$, in terms of the ungauged and tracefull perturbation $h_{\mu\nu}$, $\bar h_{\mu\nu}(h)=\hat\varPi\left(h_{\mu\nu}-2\,\nabla_{(\mu}
(\Box+\Lambda)^{-1}\chi_{\nu)}(h)\right)$. Then we apply to $S_{(2)}[\,\bar h(h)\,]$ the variational derivative with respect to $h_{\mu\nu}$ followed by the retardation prescription and, thus, finally arrive at linearized effective equations of motion \cite{serendipity},
    \begin{eqnarray}
    \left.\frac4{M_{\rm eff}^2}g^{-1/2}\frac{\delta S_{(2)}}{\delta h^{\mu\nu}}\right|_{\;++++\,\;
    \to\;-+++}^{\;\rm retarded}
    =&&\left(-\Box+\frac23\,\Lambda\right) h_{\mu\nu}
    +\frac12\,g_{\mu\nu}\left(\Box+\frac23\,\Lambda\right)h
    \nonumber\\
    &&+\,\frac12\, g_{\mu\nu}R_{(1)}
    +2\,\nabla_{(\mu}\varPhi_{\nu)}
    -g_{\mu\nu}\nabla_\alpha
    \varPhi^\alpha=0.                  \label{freequation}
    \end{eqnarray}
Here $\varPhi^\mu$ is the nonlocal function,
    \begin{eqnarray}
    &&\varPhi^\mu=\chi^\mu-\frac12\,\nabla^\mu
    \frac1{\Box+2\Lambda}\,
    \Big|_{\;\rm ret} R_{(1)},\quad
    R_{(1)}
    \equiv\nabla_\mu\chi^\mu
    -\frac12\,(\Box+2\Lambda)\,h,          \label{varPhi}
    \end{eqnarray}
whose nonlocality is given by the retarded Green's function, and $R_{(1)}$ is the linearized Ricci scalar. Now, this integro-differential equation holds, in accordance with conclusions of \cite{HiguchiMarolfMorrison,Tanaka}, in an open chart of the perturbed de Sitter spacetime and requires initial conditions at its past infinity. Remarkably, a part of this initial data follows from the equations themselves. Indeed, the trace of (\ref{freequation}) gives
    \begin{eqnarray}
    R_{(1)}
    -\frac{2\Lambda}{\Box+2\Lambda}\,
    \Big|_{\;\rm ret} R_{(1)}=0,     \label{R1equation}
    \end{eqnarray}
and this yields not only the homogeneous differential equation for $R_{(1)}$,  $\Box R_{(1)}=0$ (obtained by acting with $\Box+2\Lambda$), but also its zero initial conditions at past infinity because the second term in (\ref{R1equation}) vanishes there in view of the retarded nature of the Green's function. Therefore the linearized Ricci scalar $R_{(1)}$ of the free propagating wave is vanishing throughout the entire spacetime, $R_{(1)}(x)=0$. As a result the nonlocal function (\ref{varPhi}) coincides with the local DeWitt gauge condition function, $\varPhi^\mu=\chi^\mu$, and Eq.(\ref{freequation}) becomes absolutely identical with the linearized Einstein equations on the (A)dS background.

The rest is a typical counting of physical degrees of freedom of a propagating gravitational wave on a curved background. In the DeWitt gauge (\ref{DWgauge}) the equation $R_{(1)}(x)=0$ reduces to $(\Box+2\Lambda)h=0$. Similarly to the Feynman gauge in electrodynamics, in this gauge all components of $h_{\mu\nu}$ are propagating, but their gauge ambiguity is not completely fixed and admits residual gauge transformations $h_{\mu\nu}\to h_{\mu\nu}^{\rm phys}=h_{\mu\nu}+\nabla_\mu f_\nu+\nabla_\nu f_\mu$ with the parameter $f_\mu$ satisfying the equation
    \begin{eqnarray}
    (\Box+\Lambda)f_\mu=0.    \label{residualdiffequation}
    \end{eqnarray}
By the usual procedure these transformations can be used to select two polarizations -- free physical modes $h_{\mu\nu}^{\rm phys}=h_{\mu\nu}+\nabla_\mu f_\nu+\nabla_\nu f_\mu$. In particular, they can nullify initial conditions for $h^{\rm phys}$ on any Cauchy surface $\Sigma$ (both $\nabla_\mu f^\mu|_{\,\Sigma}$ and $\partial_0\nabla_\mu f^\mu|_{\,\Sigma}$ can be chosen to provide zero initial data for $h^{\rm phys}=h+2\nabla_\mu f^\mu$ on $\Sigma$), so that this trace identically vanishes in view of the homogeneous equation $(\Box+2\Lambda)h^{\rm phys}=0$ and makes the physical modes transverse and traceless as in the Einstein theory with a $\Lambda$-term,
    \begin{eqnarray}
    \nabla^\nu h_{\mu\nu}^{\rm phys}=0,\quad
    h^{\rm phys}=0.
    \end{eqnarray}
The remaining three pairs of initial data for $f^\mu$ accomplishes the counting of the physical degrees of freedom among spatial components of $h_{\mu\nu}$, $6-1-3=2$, while the four lapse and shift functions $h_{0\mu}$, as usual, express via the constraint equations of motion $\delta S_{(2)}/\delta h_{0\mu}=0$.\footnote{Or equivalently, when they are treated as propagating modes subject to second order in time differential equations, their initial data express via $\chi^\mu|_{\,\Sigma}=0$ and $\partial_0\chi^\mu|_{\,\Sigma}=0$.}

A similar derivation for the equation (\ref{freequation}) with a stress tensor source $2T_{\mu\nu}/M_{\rm eff}^2$ on the right hand side gives the expression for the retarded gravitational potential of a compact matter source. In the {\em DeWitt gauge} it takes the form
    \begin{eqnarray}
    &&h_{\mu\nu}
    =\frac{16\pi G_{\rm eff}}{-\Box+\frac23\Lambda}\,\Big|_{\;\rm ret}\Big(\,T_{\mu\nu}
    +g_{\mu\nu}\,
    \frac{\Box-2\Lambda}{\Box+2\Lambda}\,
    \frac{\Lambda}{3\Box}\,T \Big)
    -\nabla_\mu\nabla_\nu \frac{16\pi G_{\rm eff}}{(\Box+2\Lambda)\,\Box}\,
    \Big|_{\;\rm ret}\!T.     \label{matpot}
    \end{eqnarray}
Here the last term represents a pure gauge transformation and $G_{\rm eff}\equiv 1/8\pi M_{\rm eff}^2$ is the effective gravitational constant vs the Newton one $G_N=1/8\pi M_P^2$,
    \begin{eqnarray}
    G_{\rm eff}=\frac{\alpha(1-\alpha)}
    {A^2-\alpha^2}\,G_N.                 \label{Geff}
    \end{eqnarray}

This result was interpreted in \cite{serendipity} as a dark matter simulation -- $O(1/|\alpha|)$ amplification of the gravitational attraction due to the replacement of the Newton gravitational constant $G_N$ by $G_{\rm eff}\sim G_N/|\alpha|$ with $|\alpha|\ll 1$. This necessarily happens in the domain of positive $M^2_{\rm eff}$ with a negative $\alpha$, $\alpha<A<-\alpha$, where the factor $\alpha/(A^2-\alpha^2)\geq 1/4|\alpha|$ and $G_{\rm eff}\geq G_N/4|\alpha|\gg G_N$. For a positive $\alpha$ the theory also has $G_{\rm eff}>G_N$ for $|A|>\sqrt\alpha$. Unfortunately, however, this interpretation turned out to be misleading because the gravitational potential (\ref{matpot}) in the short ($\Box\gg \varLambda$) and long distance ($\Box\ll \varLambda$) limits does not interpolate between the flat space theory with the gravitational coupling (\ref{M_Prenorm}) and the de Sitter phase with $G_{\rm eff}$. In fact with $\Box$ ranging in these limits basically only the tensor law of gravitational coupling changes from the UV behavior
    \begin{eqnarray}
    &&h_{\mu\nu}
    \simeq -16\pi G_{\rm eff}\,\frac1{\Box_{\rm ret}}\,
    T_{\mu\nu}
    +{\rm gauge\;\;transformation},
    \quad \Box\gg\varLambda,     \label{matpotUV}
    \end{eqnarray}
to a kind of scalar gravity mediated only by a conformal mode in the IR domain
    \begin{eqnarray}
    h_{\mu\nu}
    \simeq -8\pi G_{\rm eff}\,g_{\mu\nu}\frac1{\Box_{\;\rm ret}}\,
    T+{\rm gauge\;\;transformation},
    \quad \Box\ll\varLambda     \label{matpotIR}
    \end{eqnarray}
(here we disregard details of gauge transformation terms $\sim\nabla_\mu\nabla_\nu(...)$). For any however small value of $\alpha$ the both expressions differ from the GR analogue by the tensor structure -- the gauge independent part does not coincide with the GR expression proportional to $T_{\mu\nu}-\frac12g_{\mu\nu}T$.

\subsection{Recovery of the GR limit: the physical conformal frame}
\hspace{\parindent}Breakdown of the general relativistic law can be corrected by the assumption that the physical metric $\tilde g_{\mu\nu}$ (observable and directly coupled to matter fields $\phi$) differs from $g_{\mu\nu}$ by the nonlocal conformal factor. Thus the matter action $S_{\rm matter}[\phi,\tilde g\,]$ is included into the action of the total system as
    \begin{eqnarray}
    &&S_{\rm total}[\,g,\phi\,]=S[\,g\,]+S_{\rm matter}[\phi,\tilde g[\,g\,]\,], \label{tildeaction}\\
    &&\tilde g_{\mu\nu}=
    \exp\left(\frac12\frac1{\Box
    -\mu^2}R\right)\,g_{\mu\nu},    \label{tildeg}
    \end{eqnarray}
where $\mu^2$ is some mass parameter playing the role of the potential terms $\hat P$ regulating the limit of (A)dS background. The linear perturbation of the physical frame metric (in the DeWitt gauge for $h_{\mu\nu}$) reads
    \begin{eqnarray}
    \delta\tilde g_{\mu\nu}\equiv\tilde h_{\mu\nu}= e^{-2\varLambda/\mu^2} \left(h_{\mu\nu}-\frac14\,g_{\mu\nu}
    \frac{\Box+2\varLambda}{\Box-\mu^2}\,h\right),
    \end{eqnarray}
which for short wavelengths $(\Box\gg\varLambda,\;\mu^2)$ reduces to $\tilde h_{\mu\nu}\simeq e^{-2\varLambda/\mu^2}(h_{\mu\nu}-\frac14\,g_{\mu\nu} h)$. Then the retarded potential (\ref{matpotUV}) for $\tilde  h_{\mu\nu}$ takes the GR form
    \begin{eqnarray}
    &&\tilde h_{\mu\nu}\simeq-\frac{16\pi \tilde G_{\rm eff}}{\tilde\Box}\,\Big|_{\;\rm ret}
    \Big(\,\tilde T_{\mu\nu}
    -\frac12\,\tilde g_{\mu\nu}\,\tilde T\,\Big)
    +{\rm gauge\;\;transformation},
    \quad \Box\gg\varLambda,        \label{tildematpotUV}\\
    &&\tilde G_{\rm eff}=
    e^{-2\varLambda/\mu^2}G_{\rm eff}.   \label{tildeG}
    \end{eqnarray}
where $\tilde\Box$ is based on the background de Sitter metric $\tilde g_{\mu\nu}=e^{-2\varLambda/\mu^2}g_{\mu\nu}$ and the stress tensor is defined by varying (\ref{tildeaction}) with respect to the physical metric
    \begin{eqnarray}
    \tilde T^{\mu\nu}=\frac2{\tilde g^{1/2}}\,\frac{\delta S_{\rm matter}}{\delta\tilde g_{\mu\nu}},\quad
    \tilde T_{\mu\nu}=\tilde g_{\mu\alpha}\tilde g_{\nu\beta}\,
    \tilde T^{\alpha\beta}.              \label{tildestress}
    \end{eqnarray}
A similar transition to the physical metric in the IR domain retains a purely scalar type of the gravitational potential
    \begin{eqnarray}
    \tilde h_{\mu\nu}
    \simeq -8\pi \tilde G_{\rm eff}\,\frac{\mu^2+2\varLambda}{\mu^2}\,
    \tilde g_{\mu\nu}\frac1{\tilde\Box_{\;\rm ret}}\,
    \tilde T+{\rm gauge\;\;transformation},
    \quad \Box\ll\varLambda,\mu^2.     \label{tildematpotIR}
    \end{eqnarray}

In the physical frame the general relativistic law (\ref{tildematpotUV}) is recovered with the modified value of the effective gravitational constant (\ref{tildeG}), but this is reached by the price of introducing an additional scale $\mu^2$, which of course contradicts the motivation for our model -- replacement of the numerical scale by a dynamical variable. One might think that this extra scale can be replaced by a dynamical quantity like curvature. However, by the requirement of covariance it can only be the curvature scalar, $\mu^2\to\xi R$ with some constant $\xi$, and in view of compactness of the Euclidean section of the spacetime the conformal factor in (\ref{tildeg}) becomes a numerical constant because
    \begin{eqnarray}
    \frac1{\Box-\xi R}\,R
    \equiv-\frac1\xi .                     \label{equation1}
    \end{eqnarray}
Similarly to the derivation of (\ref{equation3}) this relation holds for a compact spacetime without a boundary from the following chain of relations
    \begin{eqnarray}
    &&\frac1{\Box-\xi R}R=-\frac1\xi
    \frac1{\Box-\xi R}
   (\overrightarrow{\Box}-\xi R)1
    =-\frac1\xi
    \frac1{\Box-\xi R}
    (\overleftarrow{\Box}-\xi R)1
    =-\frac1\xi.                      \label{equation2}
    \end{eqnarray}

Thus, in contrast to anticipations of \cite{serendipity}, the perturbation theory in the original conformal frame has no GR limit either in the short wavelengths regime $\nabla\nabla\gg R$ or in the limit of $\alpha\to 0$. The failure of the correspondence principle with GR can be traced back to the level of full nonlinear equations of motion. Using (\ref{Omega}) in (\ref{eom}) one can see that in the UV limit $\nabla\nabla\gg R$ the variational derivative of the action
    \begin{eqnarray}
    &&\frac{\delta S}{\delta g_{\mu\nu}}\simeq
    \frac{M^2_{\rm eff}}2
    \,g^{1/2}\left(R_{\mu\nu}
    -\frac12\,\nabla_\mu\nabla_\nu\frac1\Box R\right)
    +O[\,E^2\,]
    \end{eqnarray}
remains nonlocal and differs from the general relativistic expression even for $\alpha\to 0$. In particular, in the approximation linear in the curvatures matter sources are coupled to gravity according to
    \begin{eqnarray}
    R_{\mu\nu}
    -\frac12\,\nabla_\mu\nabla_\nu\frac1\Box R
    +O[\,R^2\,]
    =\frac1{M^2_{\rm eff}}\,T_{\mu\nu}, \label{mattersource}
    \end{eqnarray}
where the nonlinear curvature terms $O[\,R^2\,]$ include nonlinearity in $E_{\mu\nu}$. The local Ricci scalar term of the Einstein tensor is replaced here with the nonlocal expression which guarantees in this approximation the stress tensor conservation, but contradicts the GR phase of the theory.

Absence of the GR phase, that was first noted in \cite{Solodukhin}, might seem paradoxical because the original action (\ref{action}) obviously reduces to the Einstein one in the limit $\alpha\to 0$. The explanation of this paradox consists in the observation that the transition from (\ref{action}) to the representation (\ref{newrep}) is based on the identity (\ref{equation1}) which is not analytic both in $\xi=\alpha/4$ and in the curvature. The source of this property is the constant zero mode of the scalar operator $\Box$ on compact Euclidean spacetimes without a boundary. On such manifolds the left hand side of (\ref{equation1}) is not well defined for $\xi=0$. The equivalence of the actions (\ref{action}) and (\ref{newrep}) holds only on this class of Euclidean manifolds which are motivated by the Euclidean version of the Schwinger-Keldysh technique discussed above.

In contrast to this class of manifolds, the representations (\ref{action}) and (\ref{newrep}) are not equivalent in asymptotically flat (AF) spacetime because equations (\ref{equation3}) and (\ref{equation1}) do not apply there. First, with zero boundary conditions at infinity the scalar $\Box$ does not have zero modes. Second, due to the natural AF falloff conditions, $R(x)\sim 1/|x|^4$ and $(1/\Box)\delta(x-y)\sim 1/|x-y|^2$, integration by parts in the chain of transformations (\ref{equation2}) gives a finite surface term at infinity $|x-y|\to\infty$. This leads to an alternative equation
    \begin{eqnarray}
    \frac1{\Box-\xi\,R}\,R\,\Big|_{\,\rm AF}
    =O\,[\,R\,]                      \label{equation11}
    \end{eqnarray}
with a nontrivial right hand side analytic in $\xi$ and tending to zero for a vanishing scalar curvature. This explains why the model (\ref{action}) on AF background has a good GR limit with nonlinear curvature corrections controlled by a small $\alpha$ \cite{covnonloc,serendipity}.

To recover the GR limit and, thus, the utility of the model (\ref{action}) as a possible dark energy mechanism we can again use the transition to the physical metric frame (\ref{tildeaction})-(\ref{tildeg}). It is established by a nonlocal conformal transformation , $\tilde g_{\mu\nu}[\,g\,]=e^{2\sigma[\,g\,]}\,g_{\mu\nu}$, with $\sigma=\frac14\,(\Box-\mu^2)^{-1}R$. In the UV limit this function is small, $\sigma\ll 1$, but has $O(1)$ second order derivatives, $\nabla\nabla\sigma\sim R$, so that the Einstein tensor of the physical metric $\tilde G_{\mu\nu}$ reads in terms of the original metric as
    \begin{eqnarray}
    &&\tilde G_{\mu\nu}=G_{\mu\nu}+2\big(g_{\mu\nu}\Box\sigma
    -\nabla_\mu\nabla_\nu\sigma\big)+g_{\mu\nu}\sigma_\alpha^2
    +2\sigma_\mu\sigma_\nu\nonumber\\
    &&\qquad\qquad=R_{\mu\nu}
    -\frac12\,\nabla_\mu\nabla_\nu\frac1\Box R+O\left[\,\Big(\nabla\frac1\Box R\Big)^2\right],
    \quad \sigma_\mu\equiv\nabla_\mu\sigma.
    \end{eqnarray}
We see that $\tilde G_{\mu\nu}$ in this limit in fact reproduces the left hand side of (\ref{mattersource}). Therefore, if we couple matter to the physical metric $\tilde g_{\mu\nu}$ as in (\ref{tildeaction}), then for $\tilde g_{\mu\nu}$ we will recover the usual Einstein equations
    \begin{eqnarray}
    \tilde R_{\mu\nu}
    -\frac12\,\tilde g_{\mu\nu}\tilde R
    \simeq 8\pi \tilde G_{\rm eff}\,\tilde T_{\mu\nu},
     \label{modeq}
    \end{eqnarray}
with the physical frame stress tensor (\ref{tildestress}).

Thus we get a GR phase in the conformally related frame of the theory. In the short distance regime it has in the leading order the GR retarded potential (\ref{matpotUV}), whereas for horizon and superhorizon scales it features the interaction mediated by a purely conformal mode (\ref{tildematpotIR}). Unfortunately, however, the magnitude of corrections to the GR behavior is no longer controlled by a small parameter $\alpha$ that was initially designed in \cite{serendipity} to moderate the effect of nonlocal corrections to the Einstein theory. Moreover, we could not help using an extra scale $\mu^2$ necessary for the definition of the physical metric frame (\ref{tildeg}), though this contradicts the spirit of the model motivated by the attempt not to incorporate the horizon scale (or other dimensional scales) in the action of the theory.

All this makes application of the model in realistic cosmology somewhat questionable. Nevertheless, it might be interesting as a nonlocal generalization of critical gravity theories \cite{critical} which recently became popular as holographic duals of the logarithmic conformal models. In fact, the relation (\ref{relation}) can be regarded as the analogue of the criticality condition in the local models quadratic in the curvature. It eliminates massive gravitons and for $a\neq 2$ (breakdown of the unitarity condition (\ref{arelation})) gives rise to logarithmic modes \cite{critical} corresponding to the double pole in the propagator.

Another interesting field of applications is black hole thermodynamics. They are possible due to extension of the theory from maximally symmetric to generic Einstein spaces and black hole solutions \cite{Solodukhin,newform} as stable backgrounds. In particular, as advocated in \cite{Solodukhin}, the theory (\ref{newrep}) has Schwarzschild-de Sitter black hole solutions with zero entropy in accordance with the existence of black holes of zero entropy and energy in critical gravity theories of \cite{critical}.

\section{Conclusions}
\hspace{\parindent}Several essays on nonlocal aspects of quantum field theory, gravity and cosmology that we presented here are culminating in a nonlocal cosmological model called for explanation of dark energy phenomenon. Though they look somewhat disjoint, we hope that they are strongly intertwined by the rules of handling the boundary conditions for nonlocal operations, based on the physical setup in the quantum domain. These rules are represented by the Euclidean version of the Schwinger-Keldysh technique which allows one to start with the Euclidean asymptotically flat or closed compact spacetime (much easier and universal from calculational viewpoint) and then make the transition to the Lorentzian signature setup with the retardation rule. In fact they underlie the ``integration by parts trick" of \cite{nonloccosm,DeffWood} and serve as an antithesis of the widespread, but in our opinion misleading, approach which rejects the variational nature of equations of motion in nonlocal gravity \cite{ParkDodelson}.

The efficiency of this technique was demonstrated by derivation of massless graviton modes and retarded potentials on the de Sitter background of a special nonlocal cosmology model. This ghost free model was motivated by the idea that the DE scale is not a parameter with the prefixed numerical value, but a dynamical quantity to be fixed by the mechanism of breaking the dilatation symmetry (\ref{scaling}) -- free parameter of the background solution of equations of motion. As the result this theory was shown to interpolate in a special conformal frame between a general relativistic limit and the superhorizon phase with the interaction (\ref{tildematpotIR}) mediated by the gravitational conformal mode. Though this model suffers from certain conceptual drawbacks, its cosmological implications deserve further studies, not to say that it might be interesting within the scope of the so called critical gravity theories \cite{critical}.

Nonperturbative results of Sect.2.3 for the effective action, based on the nonlocal late time asymptotics of the heat kernel, are standing somewhat apart from direct applications in gravity and cosmology. Rather exotic nonlinear and nonlocal structures (\ref{4.8})-(\ref{4.11}) describe the modification of the Coleman-Weinberg potential caused by the transition between the compact domain of nearly constant field to its zero value at spacetime infinity. Thus they might find implications in quantum back reaction problems of black hole thermodynamics and deserve generalization to asymptotically (A)dS-spaceitmes to serve as one more source of a nonlocal effective modification of the Einstein theory. All this makes the class of nonlocal models of the above type open for interesting future studies.

\section*{Acknowledgements}
This paper is based on the series of results obtained in the course of many years of collaboration with G.A.Vilkovisky, and I am deeply grateful to him for a moving spirit behind the strategy and motivation for this work. The author also strongly benefitted from fruitful discussions and correspondence with S. Deser, S. Solodukhin and R. Woodard.  This work was partly supported by the RFBR grant No. 14-02-01173.

\appendix
\renewcommand{\theequation}{\Alph{section}.\arabic{equation}}

\section{Surface terms in nonlocal gravity: Schwarzschild-de Sitter background}
Transition to the new representation of the nonlocal gravity is based on the identity (\ref{equation1}) which holds for a compact spacetime without a boundary or under boundary conditions which do not generate surface terms under integration by parts in (\ref{equation2}). Here we check this property for the conical singularity arising, for example, at the cosmological horizon of the Schwarzschild-de Sitter Euclidean metric.

Close to the conical singularity the metric behaves as
    \begin{eqnarray}
    ds^2\simeq d\rho^2+\varkappa^2\rho^2\,d\phi^2+R^2_{\rm hor} d\Omega^2,\quad\rho\to 0,
    \end{eqnarray}
where $\rho$ and $\phi$ are the relevant radial and angular coordinates, $R_{\rm hor}$ is a size of the horizon spanned by the rest of angular variables $\Omega$, and $\varkappa\neq 1$ characterizes the deficit angle. The corresponding wave equation for a scalar field has a leading contribution for the $m$-th angular harmonics
    \begin{eqnarray}
    &&\left[\,\frac{d^2}{d\rho^2}+\frac1\rho\,\frac{d}{d\rho}-
    \frac{m^2}{\varkappa^2\rho^2}-\frac{l(l+1)}{R^2_{\rm hor}}\,\right]\varphi(\rho,\phi,\Omega)=0,\\
    &&\nonumber\\
    &&\varphi(\rho,\phi,\Omega)=\rho^\alpha\,e^{im\phi}\,\psi_m(\Omega),
    \end{eqnarray}
where the power $\alpha$ should be determined from this equation.

At $\rho\to 0$ it reduces to the quadratic equation for $\alpha$
    \begin{eqnarray}
    \left(\,\alpha(\alpha-1)+\alpha-
    \frac{m^2}{\varkappa^2}\,\right)\,\frac1{\rho^2}
    +O\left(\frac1\rho\right)=0
    \end{eqnarray}
or
    \begin{eqnarray}
    \alpha=\pm\frac{m}{\varkappa}.
    \end{eqnarray}

Thus the two linear independent solutions read
    \begin{eqnarray}
    \varphi_\pm=\rho^{\pm m/\varkappa}\,
    e^{im\phi}\,\psi_m(\Omega),\quad m\neq 0
    \end{eqnarray}
and for the case of coincident roots with $m=0$ there are also two asymptotic solutions -- a constant and logarithmic ones
    \begin{eqnarray}
    \varphi_\pm=\psi_0(\Omega),\quad \ln\rho\;\overline\psi_0(\Omega).
    \end{eqnarray}
The corresponding Green's function has for the most general choice of boundary conditions all these asymptotic behaviors
    \begin{eqnarray}
    G(x,y)\,|_{\;\rho_x\to 0}\sim\varphi_\pm,
    \end{eqnarray}
and all of them {\em except the logarithmic part} give a vanishing surface term in (\ref{equation2})
    \begin{eqnarray}
    &&\int\limits_{\rho_x\to 0} dS_x^\mu\,g^{1/2}\,1\,
    \overleftrightarrow{\partial_\mu} G(x,y)=\int_0^{2\pi} d\phi\,\int d\Omega\,\varkappa\,\rho\,
    \partial_\rho G_\pm\Big|_{\;\rho\to 0}\nonumber\\
    &&\qquad\qquad\qquad=\pm m\,\rho^{\pm m/\varkappa}\,\delta_{m0}\int d\Omega\,G_m(\Omega,y)\Big|_{\;\rho\to 0}=0.
    \end{eqnarray}
On the contrary, the log part of the Green's function gives
    \begin{eqnarray}
    &&\int\limits_{\rho_x\to 0} dS_x^\mu\,g^{1/2}\,1\,
    \overleftrightarrow{\partial_\mu}\,G(x,y)=
    \varkappa \int d\Omega\, G_0(\Omega)\neq 0.
    \end{eqnarray}

Usually the power $G\sim \rho^{-m/\varkappa}$ and logarithmic $G\sim \ln\rho$ singularities are forbidden by boundary conditions. Then the conical singularity is harmless in (\ref{equation2}) for all regular boundary conditions.

\end{document}